\documentclass{aa}
\usepackage[dvips]{graphicx}
\newcommand{\asec}{$^{\prime\prime}$}

\def\AMM{NH$_3$}
\def\ace{CH$_{3}$C$_{2}$H}

\def\CIII{\mbox{C$^{17}$O}}

\def\HII{H{\sc ii}}

\def\kms{\mbox{km~s$^{-1}$}}
\def\cmc{cm$^{-3}$}
\def\cmq{cm$^{-2}$}

\begin{document}
\title{Search for massive protostellar candidates in the southern hemisphere: 
I. Association with dense gas
\thanks{Based on results collected at the European Southern Observatory
(ESO), La Silla, Chile.}}
\author{F. Fontani \inst{1} \and 
	M.T. Beltr\'an \inst{2} \and J. Brand \inst{3}
	\and R. Cesaroni \inst{2}
	\and L. Testi \inst{2} 
	\and S. Molinari \inst{4} \and C.M. Walmsley \inst {2}
        }
\institute{Dipartimento di Astronomia e Fisica dello spazio, Largo E. Fermi 2,
           I-50125 Firenze, Italy \and
	   INAF, Osservatorio Astrofisico di Arcetri, Largo E. Fermi 5,
           I-50125 Firenze, Italy \and
	   Istituto di Radioastronomia, CNR, Via Gobetti 101, I-40129 Bologna,
	   Italy \and
	   IFSI, CNR, Via Fosso del Cavaliere, I-00133 Roma, Italy
 }
\offprints{F. Fontani, \email{fontani@arcetri.astro.it}}
\date{Received date; accepted date}

\titlerunning{massive protostars in the southern hemisphere}
\authorrunning{Fontani et al.}

\abstract{We have observed two rotational transitions of both CS
and \CIII , and the 1.2~mm continuum emission towards
a sample of 130 high-mass protostellar candidates with $\delta<-30^{\circ}$. 
This work represents the first step of the extension to the southern 
hemisphere of a project started more than a decade ago aimed at the 
identification of massive protostellar candidates.
Following the same approach adopted for sources with 
$\delta\geq-30^{\circ}$, we have selected from the IRAS Point Source 
Catalogue 429 sources which potentially are compact molecular clouds
on the basis of their IR colours. The sample has then been divided into two 
groups according to the colour indices [25$-$12] and [60$-$12]: the
298 sources with [25$-$12]$\geq$0.57 and [60$-$12]$\geq$1.30 have been
called {\it High} sources, the remaining 131 have been called
{\it Low} sources. In this paper, we check the association
with dense gas and dust in 130 {\it Low} sources. We have obtained 
a detection rate of $\sim 85\%$ in CS, demonstrating a tight 
association of the sources with dense molecular clumps.
Among the sources detected in CS,
$\sim 76\%$ have also been detected in
\CIII\  and $\sim 93\%$ in the 1.2~mm continuum.
Millimeter-continuum maps show the presence of clumps with diameters
in the range $0.2 - 2$ pc and masses from a few $M_{\odot}$ to 
$10^{5}M_{\odot}$; H$_{2}$ volume densities 
computed from CS line ratios lie between $\sim 10^{4.5}$ and $10^{5.5}$\cmc .
The bolometric luminosities of the sources, derived from IRAS data, are in the
range $10^{3}-10^{6}\;L_{\odot}$, consistent with embedded high-mass objects.
Based on our results
and those found in the literature for other samples
of high-mass young stellar objects, we conclude that our sources are
massive objects in a very early evolutionary stage, probably
prior to the formation of an \HII\ region. We
propose a scenario in which {\it High} and {\it Low} sources are
both made of a massive clump hosting a high-mass protostellar
candidate and a nearby stellar cluster. The difference might be due 
to the fact that the 12 $\mu$m
IRAS flux, the best discriminant between the two groups, is dominated by
the emission from the cluster in {\it Lows} and from the massive 
protostellar object in {\it Highs}.
}

\maketitle

\section{Introduction}
\label{intro}

Recently, an ever growing effort has been devoted to investigating
the early evolutionary stages of massive stars ($M\geq 8 M_{\odot}$).
In particular, attention has gradually shifted from the study of newly
formed ZAMS stars to
objects in an earlier evolutionary stage, prior to the formation of
an \HII\ region, deriving their luminosity from the release of 
gravitational energy: these objects are named {\it protostars}.
The observational approach to searching for high-mass protostars was first 
formulated by Habing \&
Israel (\cite{habing}): likely candidates must be associated 
with dense circumstellar environments, not be associated with HII regions,
and they should have high luminosities.

Following these criteria, with the aim of identifying massive protostellar 
candidates (with $\delta\geq-30^{\circ}$), 
Palla et al. (\cite{palla}) selected a
sample of 260 sources from the IRAS Point Source Catalogue (IRAS-PSC) 
with 60$\mu$m flux greater than 100~Jy and colours satisfying the 
criteria established by Richards et al. (\cite{richards}) for
compact molecular cores. This sample was then divided into two groups
according to their [25$-$12] and [60$-$12] colours: the {\it High} sources, 
which have [25$-$12]$\geq$0.57 and [60$-$12]$\geq$1.30 characteristic of association with UC \HII\ regions 
(Wood \& Churchwell~\cite{wec}), and
the {\it Low} sources, with complementary colours. Palla et al.
found a lower association
rate with H$_2$O masers for the {\it Low} sources, and interpreted
this as an indication of relative youth.
In order to confirm this result, and to better understand the nature of 
{\it High} and {\it Low} sources, the whole sample has been studied in various 
tracers, including
molecular lines and continuum emission, from centimeter to near-infrared
wavelengths (Molinari et al. \cite{mol96}, \cite{mol98a}, \cite{mol00}, 
\cite{mol02}; Brand et al. \cite{brand}; Zhang et al.~\cite{zhang}). The main findings of these studies
are the following:
\begin{itemize}
\item {\it High} and {\it Low} sources have luminosities typical of 
intermediate- or high-mass objects ($L\geq10^{3}L_{\odot}$);
\item a relatively large fraction of {\it High} sources is associated with 
UC \HII\ regions ($57\%$, Molinari et al.~1998a);
\item the {\it Low} group contains a considerable fraction ($\sim 76\%$)
of likely precursors of stars with mass $M>10M_{\odot}$.
\end{itemize}
Furthermore, Molinari et al. (\cite{mol98b}) and more recently Fontani et al. 
(\cite{fonta1}, \cite{fonta2}) have studied in detail, at low
and high angular resolution, three
sources belonging to the {\it Low} group which have been proposed as
protostellar candidates: IRAS 23385+6053, IRAS 21307+5049 and
IRAS 22172+5549. In all three cases, they have detected a compact
($\sim 0.03-0.04$ pc), dense ($\sim 10^{7}$\cmc ) and massive
($\sim 50-300 M_{\odot}$)
molecular envelope, likely hosting an intermediate- to high-mass
YSO in the protostellar phase. 

The results obtained for sources with $\delta\geq-30^{\circ}$
suggested an extension to sources with $\delta <-30^{\circ}$
following the same approach, in order to complete this study. 
With this motivation, we have applied 
the selection criteria of Palla et al.~(\cite{palla}) to
sources of the IRAS-PSC with $\delta <-30^{\circ}$, finding
298 {\it High} and 131 {\it Low} sources. It is worth noting that
the samples selected by us likely contain a higher contamination 
of \HII\ regions than those selected by 
Palla et al.~(\cite{palla}), because surveys of \HII\ regions south of 
$\delta <-30^{\circ}$ are much less numerous than those 
with $\delta\geq-30^{\circ}$.

When identifying massive protostellar candidates, the first step is to
establish an association with dense molecular clumps. 
Dense gas is traced by warm dust emission from a massive core 
at millimeter and sub-millimeter wavelengths, 
and by millimeter rotational and inversion
transitions of various molecular species, such as CS, \AMM\ and \CIII . 
In this paper we present 
observations obtained with the SEST-15m telescope of rotational 
transitions of CS and \CIII , and of the 1.2~mm 
continuum emission towards almost all (130 out of 131)
sources belonging to the {\it Low} subsample. 

All sources belonging to the {\it High} subsample
have already been observed in the CS (2$-$1) line 
by Bronfman et al. (\cite{bronfman}), who performed a
complete survey in this line towards IRAS sources with [25$-$12]
and [60$-$12] colours characteristic of UC \HII\ regions with the
SEST and the Onsala telescope. 

Also, an alternative sample of high-mass protostellar candidates has been 
selected from the IRAS-PSC by Sridharan et al.~(\cite{sridharan}). They 
used selection criteria similar to those
of Palla et al.~(\cite{palla}), with the important difference that 
they ruled out all {\it Low} sources, and therefore their sample is basically
made of {\it High} sources. Then,
Beuther et al.~(\cite{beuther}) observed the molecular environment
associated with these sources in some CS lines and 1.2~mm continuum.
Therefore, their results, as well as those of 
Bronfman et al.~(\cite{bronfman}),
are of great interest for the present paper, and will be
used in the following for the sake of comparison to our findings.
Hereafter, the sample selected by Sridharan et al.~(\cite{sridharan})
and observed in various tracers by Beuther et al.~(\cite{beuther})
will be called ``Sridharan/Beuther sample''.
All the sources detected in the CS 
(2$-$1) line by Bronfman et al.~(\cite{bronfman})
have been observed by Faundez et al.~(\cite{faundez}) in the 1.2~mm 
continuum. However, we have not used their results in this paper
becuase their work was published after our 
paper was submitted. They will be discussed in a forthcoming
paper (Beltran et al., in prep.) entirely devoted to the observations
of the millimeter continuum (see Sect.~\ref{contdata}).

Sect.~\ref{obs} describes the observations, and Sect.~\ref{res} presents
the results. In Sect.~\ref{phipar} we derive the physical 
properties of the molecular clumps, which we discuss in Sect.~\ref{discu}.
The conclusions are summarized in Sect.~\ref{conc}.

\section{Observations}
\label{obs}

\subsection{Molecular lines}

Single-pointing observations of CS and \CIII\
were obtained with the SEST (Swedish-ESO Submillimetre Telescope)
15-m telescope at ESO-La Silla, Chile. In Table~\ref{tobs} we give 
the molecular transitions
observed (Col.~1), the line rest frequencies (Col.~2), the telescope 
half-power beam width (HPBW, Col.~3), and the  
channel spacing (Col.~4) and total bandwidth (Col.~5) of the spectrometer used.

All observations were carried out towards the positions of the 
IRAS sources given in Table~\ref{tsources}. We observed using 
dual beam switching 
with a $11^{\prime}37$\asec\ throw. The data were calibrated with the chopper
wheel technique (see Kutner \& Ulich~\cite{kutner}). Pointing was checked 
every $1-2$ hours on SiO masers at 7~mm. The pointing accuracy is
estimated to be $\sim 3$\asec .
\begin{table*}
\begin{center}
\caption[] {Observed transitions}
\label{tobs}
\begin{tabular}{ccccc}
\hline \hline
 molecular  & frequency & HPBW & $\Delta v^{\dagger}$ & Bandwidth$^{\dagger}$ \\
 transition  & (GHz) & (\asec ) & (\kms ) & (\kms ) \\
\hline
CS (2$-$1)  & 97.980 & 51 & $0.13/2.12$ & $130/3053$ \\
CS (3$-$2)  & 146.969 & 34 & $0.087/1.43$ & $87/2059$ \\
CS (5$-$4)  & 244.935 & 21 & $0.052/0.86$ & $52/1238$ \\
\CIII\ (1$-$0) & 112.359 & 45 & $0.11/1.85$ &  $110/2664$\\
\CIII\ (2$-$1) & 224.714 & 22 & $0.057/0.93$ & $57/1342$ \\
\hline
\end{tabular}
\end{center}
$^{\dagger}$ the two values refer to the two spectrometers used. 
\end{table*}

\subsubsection{CS}
\label{csobs}

Observations of the CS (2$-$1), (3$-$2) and (5$-$4) lines were performed 
from May 23 to 25, 2001, and from May 6 to 11, 2002.
We observed the (2$-$1), (3$-$2) and (5$-$4) lines respectively
in 130, 128 and 3 out of 131 sources of the initial sample.
The antenna temperature, $T^{*}_{A}$ and the main beam
brightness temperature $T_{\rm MB}$ are related as:
$T_{\rm MB}=T^{*}_{A}/\eta_{\rm MB}$, with $\eta_{\rm MB}=0.73$,
0.66 and 0.50 for CS (2$-$1), (3$-$2) and (5$-$4) respectively.

We simultaneously observed the (2--1) and (3--2) lines during the first
observing run, and the (2--1) and (5--4) lines during the second observing
run, using two Acusto-Optic Spectrometers: one with low spectral
resolution and large bandwidth, and a second with higher
spectral resolution and smaller bandwidth (see Table~\ref{tobs}).
Since the $v_{\rm LSR}$
was unknown for most sources we tuned the receivers to the $v_{\rm LSR}$
of the tangent point for the source galactic longitude 
(see Table~\ref{tsources}).
The values of $v_{\rm LSR}$ used during the first and the second observing
runs are listed in Cols.~4 and 5, respectively. 
The integration time ranged from 3 to 4.5 minutes.
For some lines detected at the edge of the bandwidth in the first scan
we integrated the minimum possible time to get 
a low S/N detection of the CS (2$-$1) line in low resolution; after having
determined the $v_{\rm LSR}$ of the line, we then re-centered the high
resolution backend and made another measurement.

For 5 sources (13558$-$6159, 15262$-$5541, 16170$-$5053, 16402$-$4943, 
16581$-$4212) observed during the first run we obtained bad quality
spectra. We thus repeated the observations in the second run. 

\subsubsection{\CIII\ }
\label{c17oobs}

\CIII\ (1$-$0) and (2$-$1) lines were observed in the period from
May 6 to 11, 2002. For sources previously detected in CS we have 
used the LSR velocity of these lines to center the backends.
We have also observed 8 objects not detected in CS.
For these, we used the same velocity adopted for the CS observations, 
computed as described in Sect.~\ref{csobs}.
We observed the (1$-$0) and (2$-$1) 
transitions simultaneously using the 3 and 1.3~mm receivers.
As for the CS lines, $T_{\rm MB}$ and $T^{*}_{A}$ are related as
$T_{\rm MB}=T^{*}_{A}/\eta_{\rm MB}$, with $\eta_{\rm MB}=0.70$ and 0.50 
for the \CIII (1$-$0) and (2$-$1) lines, respectively.

\subsubsection{\CIII\ fitting procedure}
\label{c17ofit}

The \CIII\ (1$-$0) and (2$-$1) rotational transitions have hyperfine
structure (see e.g. Frerking \& Langer \cite{frerking}). To take 
this into account, we fitted the lines using METHOD HFS  
of the CLASS program, which is part of the GAG-software developed at
the IRAM and the Observatoire de Grenoble. This fits the lines assuming 
that all components 
have equal excitation temperatures, that the line separations are 
fixed at the laboratory values, and that the line widths are identical.
This method also gives an estimate of the total
optical depth of the lines based on the intensity ratio of the 
different hyperfine components.

\subsection{Continuum}
\label{contobs}

The 1.2~mm continuum observations were carried out with the 37$-$channel
bolometer array SIMBA (SEST Imaging Bolometer Array) at the SEST, on July
16$-$20, 2002 and July 9$-$13, 2003. 

Maps were obtained towards all sources detected in CS, with
the exception of 10555$-$5949, and towards 12 sources undetected in CS.
Around all IRAS sources, we mapped a region of size 900\asec\ $\times$
400\asec\ (azimuth $\times$ elevation), which was scanned at a rate of
80\asec /s. 
The total integration time per map was about 15~minutes, and
the typical noise level in the maps is 25$-$40~mJy/beam.
Atmospheric opacity was determined from skydips, which were taken every 2
hours, and values at the zenith ranged between 0.21 and 0.50 (in 2002) and
0.13 and 0.30 (2003). The data were calibrated using observations of
Uranus, made once
or twice per day; the conversion factor ranged between 58 and 75
mJy/count in 2002, and between 50 and 69 in 2003. 
The pointing of the SEST was determined to be accurate 
within a few arcsec, by observing a strong continuum source every 2 hours.
The HPBW is $\sim 24$ \asec .

All data were reduced with the program MOPSI, written by R. Zylka
(Grenoble), and according to the instructions given in the SIMBA Observers
Handbook\footnote{http://puppis.ls.eso.org/staff/simba/manual/simba/
index.htm.}, (2003).

\section{Observational results}
\label{res}

The observed sources are listed in 
Table~\ref{tsources}. Column 1 gives the IRAS name,
and the equatorial (J2000) coordinates of the IRAS source
are listed in Cols.~2 and 3, respectively. 
In Cols.~4 and 5 we list the center velocities used for the 
CS observations during the first and the second
run, respectively, chosen as explained in Sect.~\ref{csobs}.
In Cols.~6 to 8 we present the following information: detection (Y) or
non-detection (N) in CS, \CIII\ and 1.2~mm continuum,
respectively (N.O. means ``not observed''). For the millimeter continuum, 
we have considered as detected those sources which show emission 
above the 3$\sigma$ level in the maps. 
In Col.~9 we give the angular separation, $\Delta$, between the IRAS 
source and the peak position of the millimeter continuum. 
For sources with multiple peaks (see Sect.~\ref{contdata}), $\Delta$ 
represents the separation between the IRAS position and the nearest peak.

\begin{table}
\begin{center}
\caption[] {Number of sources detected (Y), not detected (N) or 
not observed (N.O.) in \CIII\ and 1.2~mm continuum among the 111 sources 
detected in CS. }
\label{cross}
\begin{tabular}{cccc}
\hline \hline
 & Y & N & N.O. \\
\hline
\CIII\ &   84 & 26 & 1 \\
1.2~mm & 101 & 8 & 2 \\
\hline
\end{tabular}
\end{center}
\end{table}

\subsection{CS lines}
\label{csdata}

We observed 130 out of 131 sources of the initial sample, and
detected CS emission in 111 of them, with a detection
rate of $\sim 85\%$. This indicates a tight 
association of the sample with dense gas, as 
partially expected on the basis of our experience with sources
in the northern hemisphere.  
In 14 sources, we have detected only the 
CS (2$-$1) transition, while in one source, 10555$-$5949, only 
the (3$-$2) line was detected. Only 3 sources were observed in 
CS (5$-$4), but none were detected.
One can compare this result with that found by Bronfman et 
al.~(\cite{bronfman}) obtained with the SEST 
and the Onsala Telescope in their less sensitive survey 
toward {\it High} sources (see Sect.~\ref{intro}): 
they found a detection rate of $\sim 80\%$ in the sources
observed with the SEST. Adopting the same detection 
limit as Bronfman et al.~(\cite{bronfman}) ($3\sigma\simeq 0.3$ K in main beam 
brightness temperature $T_{\rm MB}$, which corresponds to
$\sim6\sigma$ in our observations) we obtain a comparable detection rate 
of $\sim 70\%$. This shows that both {\it High} and {\it Low} 
sources with $\delta<-30^{\circ}$ are similarly associated with dense gas.

In Table~\ref{tcsline} we list the CS line parameters obtained from the 
high resolution spectra. Almost all observed lines are 
well fitted by Gaussians, and the parameters
have been calculated from these fits, except where specified otherwise
(see discussion below). The integrated intensities, 
$\int T_{\rm MB}{\rm d}v$, of the CS (2$-$1) and (3$-$2) lines are given
in Cols.~2 and 5, respectively; 
the line peak velocities, $v_{\rm LSR}$, are 
listed in Cols.~3 and 6, and Cols.~4 and 7 list the line
widths at half maximum, FWHM. In Cols.~2 and 5 we also give 
the $3\sigma$ level, in K \kms , of the spectra for non-detected sources,
obtained assuming an average value of FWHM=2.5 \kms .

Several spectra show multiple
velocity components (16187$-$4932, 16254$-$4844, 16344$-$4605, 16363$-$4645,
16535$-$4300, 17036$-$4033, 17225$-$3426, 17256$-$3631, 17285$-$3346, 
17355$-$3241). For these sources, we have 
performed, where possible, Gaussian fits to all components
(the corresponding Gaussian parameters are listed in Table~\ref{tcsline}).
The two components detected in 16535$-$4300 have also been detected
in the \CIII\ lines, whereas for 16363$-$4645 none of them were
detected in \CIII . For all other sources with multiple components,
only one component has also been detected in \CIII .
In Sect.~\ref{phipar}, where we will derive the physical 
properties of the clumps, we will refer to the CS component which has also
been revealed in \CIII ; for 16363$-$4645 and 16535$-$4300 we have
chosen the strongest one.

There are cases also in which the line profiles are asymmetric or
deviate significantly from a Gaussian shape.
In some spectra the lines have two blended peaks 
(e.g. 10123$-$5727, 17040$-$3959 and 17377$-$3109). This could be due to the 
superposition of separate velocity components, 
or to self$-$absorption. Also, a few lines present broad wings 
(e.g. 13438$-$6203, 15579$-$5303, 16218$-$4931, and 16204$-$4916). 
Incidentally, we note that
15579$-$5303 and 16204$-$4916 have FWHM very much higher
than all other lines ($\sim10-11$ \kms ).
For these lines, the 
parameters listed in Table~\ref{tcsline} have been derived from moment 
integrals over the velocity intervals indicated in parentheses in 
Cols.~3 and 7.

\subsection{\CIII\ lines}
\label{c17odata}

We observed the \CIII\ (1$-$0) and (2$-$1) lines towards all of the 111
sources detected in CS, and towards 8 sources not detected
in CS. Emission was detected in 84 sources, all of them previously detected
in CS (see Table~\ref{cross}): the \CIII\ detection rate is thus $\sim76\%$ 
for sources detected in CS,
and $\sim71\%$ for all observed sources.  
In Table~\ref{tcoline} we give the line parameters obtained from the high
resolution spectra: in Cols.~3, 4, 5 and 6 we list integrated 
intensity ($\int T_{\rm MB}{\rm d}v$), peak velocity ($v_{\rm LSR}$), FWHM 
and opacity ($\tau_{10}$) of the
\CIII\ (1$-$0) line, respectively. 
Cols.~8, 9, 10 and 11 show the same
parameters for the \CIII\ (2$-$1) line.
The integrated intensities have been computed from integrals over the 
velocity ranges given in Cols.~2 and 7 of Table~\ref{tcoline}, while for the
other parameters we have adopted the fitting procedure
described in Sect.~\ref{c17ofit}. 

\begin{figure}
\centerline{\includegraphics[angle=0,width=8.cm]{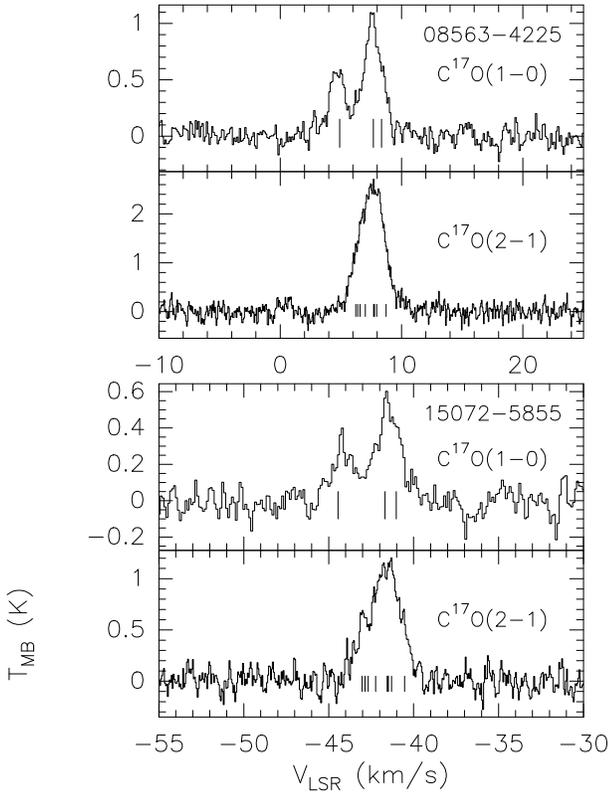}}
\caption{Spectra of the \CIII\ (1$-$0) and (2$-$1) lines 
obtained towards 08563$-$4225 (top panels) and 15072$-$5855 (bottom
panels). Vertical lines under the spectra
indicate the hyperfine components.}
\label{f08563}
\end{figure}

In several spectra the different
hyperfine components are fairly well-resolved. Two examples are shown in
Fig.~\ref{f08563}, where we have also indicated the position of the 
hyperfine components.
Only one source, 16535$-$4300, presents a secondary fainter velocity component
as in the corresponding CS spectra.  
The optical depths show that the \CIII\ (1$-$0) and (2$-$1) lines are 
optically thin in almost all detected sources. Therefore, in
Sect.~\ref{tkin} we will assume optically thin conditions when 
computing kinetic
temperatures and column densities of the molecule.

\subsection{Millimeter continuum}
\label{contdata}

We mapped 124 sources in the 1.2~mm continuum, among which 109 out of
the 111 sources detected in CS, and a further 15 sources not 
detected in CS. Since the present paper is focused on the molecular emission,
hereafter we will discuss only the maps of the sources detected in CS.
The analysis of the 1.2~mm continuum maps of all observed
sources will be available in a
forthcoming paper (Beltran et al., in prep.) completely devoted to
this purpose. In that work, we will also compare our data with those of
Faundez et al.~(\cite{faundez}), who observed the 1.2~mm continuum
emission towards a sample of {\it High} sources. 

\begin{figure}
\centerline{\includegraphics[angle=0,width=7cm]{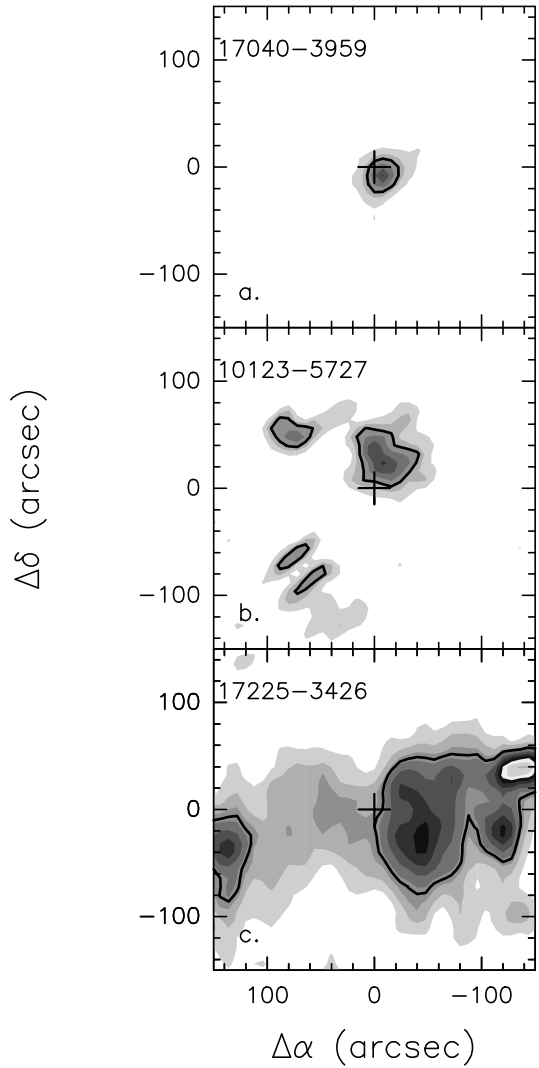}}
\caption{Examples of various morphologies of the 1.2~mm continuum
emission. {\bf a}: 1.2~mm continuum map of 17040$-$3959. Contour levels range from
0.09 ($\sim 3\sigma$) to 0.54 by 0.09 Jy beam$^{-1}$. The cross indicates the 
position of the
IRAS source. The solid line corresponds to the FWHM.
{\bf b}: same as {\bf a} for 10123$-$5727. Contour levels range
from 0.1 ($\sim 3\sigma$) to 0.7 by 0.1 Jy beam$^{-1}$. {\bf c}: same as {\bf a}
for 17225$-$3426. Contour levels range from 0.07 ($\sim 3\sigma$) to 1.07 by 
0.1 Jy beam$^{-1}$.}
\label{fcont_maps}
\end{figure}

The observations show the presence of dusty clumps in 101 out of 
109 sources previously detected in CS (see Table~\ref{cross}), 
which translates into a detection rate of $\sim 93\%$.
Morphologically, the maps show a large variety of features and structures. 
In several cases we detected an isolated clump, but only a small
fraction of the clumps show a simple 
spherical symmetry (e.g. 17040$-$3959). Most of them have an  
elongated shape (e.g. 15557$-$5337), secondary faint peaks 
(e.g. 16535$-$4300) or a core-halo structure 
(e.g. 13481$-$6124). Additionally, the majority of the maps shows 
multiple clumps. One can distinguish between sources in which the clumps 
are separable (i.e. with the contours at half of the maximum well
separated), 
and sources in which they are superimposed and not separable.
In Fig.~\ref{fcont_maps} we show an example of an isolated spherical clump,
17040$-$3959, and two examples of ``clumpy'' sources: in 10123$-$5727 the 
clumps are separable, while in 17225$-$3426 they are not.
 
As already said, a detailed analysis of these maps will be
provided in a forthcoming paper.
For the present discussion, we concentrate only on the 
continuum source associated with the line emission. 
Some physical parameters of the clumps, derived 
from the molecular line data in Sect.~\ref{phipar},
require an estimate of the source angular diameter.
Since we have not made maps in the molecular lines, this estimate
has been obtained from the continuum maps, making the assumption that 
the millimeter continuum and the molecular lines trace the same region. 
Although some authors have 
shown that in clumps associated with high-mass YSOs
dust and molecular line emission may have different distribution 
(e.g. Fontani et al.~\cite{fonta2}), typically the angular diameters 
of their emitting regions are comparable. 
Thus, the assumption that the millimeter continuum and the 
molecular lines trace the same region is a reasonable approximation. 

The lines were observed towards the position of the IRAS source;
we have thus searched for the continuum source which is closest to the 
IRAS position. 
For sources with multiple clumps we have assumed that the 
IRAS source is associated with a particular
continuum clump if the IRAS source lies within the clump's 3$\sigma$
contour level.
For sources in which none of the clumps includes the IRAS position, 
we have not assigned any ``continuum source''.

Finally, several sources require a comment:
\begin{itemize}
\item 08477$-$4359: the IRAS position is between three faint peaks.
In Table~\ref{tsources}, $\Delta$ is related to the nearest of these peaks.
\item 14131$-$6126 and 14395$-$5941: we have identified a single clump, but
the observed contour at half maximum is comparable to the HPBW. They
are hence to be considered point-like sources;
\item 16106$-$5048 and 16573$-$4214: have a filamentary structure similar to 
17225$-$3426 (Fig.~\ref{fcont_maps}), in which we could not identify
a clump coincident with the IRAS source;
\item 15038$-$5828, 16153$-$5016, 16254$-$4844, 16403$-$4614, 16417$-$4445, 16428$-$4109,
17156$-$3607: show \CIII\ emission but no continuum 
emission. However, the \CIII\ lines are faint, and in several of these
sources the non-detection of
the millimeter continuum may be a ``distance'' effect. 
We will discuss this point in Sect.~\ref{mass}.
\end{itemize}

\section{Derivation of the physical parameters}
\label{phipar}

The main physical properties of the sources are presented in three tables:
the parameters that do not depend on the source distance are listed in 
Table~\ref{tdist_ind}; in Table~\ref{tdist_dep} we give distances, 
source linear diameters and luminosities, and in Table~\ref{tmass} we 
list the mass estimates. We now 
outline the methods used to derive each parameter, and present the results
obtained.

\subsection{Distance-independent parameters}

The physical parameters which do not depend on the source distance are
listed in Table~\ref{tdist_ind}: the angular diameter
of the clumps (Col.~2), the 1.2~mm continuum flux densities (Col.~3),
the kinetic temperature, the \CIII\ column density and the H$_{2}$ total column
density of the gas derived from the \CIII\ lines (Col.~4, 5 and 6, respectively) 
and the H$_{2}$ volume density obtained from the CS data (Col.~7). 

\subsubsection{Angular diameters}
\label{ang}

The angular diameters ($\theta$) of the clumps identified 
in the 1.2~mm continuum maps (see Sect.~\ref{contdata})
have been computed assuming the sources are Gaussian, and 
deconvolving the observed contour at half maximum with
a Gaussian beam of 24\asec . No angular diameter has been
attributed to those sources for which the IRAS position does not lie
within the 3$\sigma$ contour of any of the clumps identified in the maps. 
Moreover, we could not 
compute the angular diameters of 14131$-$6126 and 14395$-$5941, because they
are not resolved (see Sect.~\ref{contdata}).
The diameters range from $\sim 10$\asec\ to $\sim 70$\asec , and are
distributed around $\sim 35$\asec\ (see Fig.~\ref{histo_t_n} (a)).

In Fig.~\ref{histo_t_n} (b) we plot the distribution of the
quantity $\Delta/\theta$, where $\Delta$ is the angular separation between 
the IRAS position and the millimeter continuum peak, listed 
in Table~\ref{tsources}. 
$\sim 80\%$ of the sources have $0\leq\Delta/\theta\leq 1$, implying
that the large majority of the identified clumps are indeed associated with
the corresponding IRAS source. The maximum nominal uncertainty in 
the IRAS position is $\sim 16$ \asec . To estimate the effect of this
error on the quantity $\Delta/\theta$, we have considered the most
pessimistic possibility by computing the distribution 
of the ratio $(\Delta+16)/\theta$: we find that the number of sources 
with $0\leq\Delta/\theta\leq 1$ reduces to $\sim 50\%$. However, it is very unlikely that all the IRAS positions are
affected by the maximum error. An "average" of 8 \asec\ is more
plausible. Therefore, we have 
derived the distribution of $(\Delta+8)/\theta$: in this case 
the fraction of sources with $0\leq\Delta/\theta\leq 1$ is $\sim 70\%$,
which is very close to that found if we neglect the position
uncertainty. Therefore, we believe that the uncertainty in the IRAS 
position do not significantly affect the distribution of $\Delta/\theta$.

Continuum flux densities, $F_{\nu}$ (Col.~3 of Table~\ref{tdist_ind}), 
are obtained by integrating the maps 
over polygons circumscribing the identified clumps. 
For isolated clumps, this polygon corresponds to 
the 3$\sigma$ contour level; for multiple clumps, we have 
determined the polygon ``by eye'', trying to cut out contributions of
secondary sources close to the main clump. 

\subsubsection{Rotation temperature and H$_{2}$ column density from \CIII\ lines} 
\label{tkin}

Cols.~4 and 5 of Table~\ref{tdist_ind} list the rotation temperatures,
$T_{\rm rot}$, and \CIII\
column densities, $N_{\rm C^{17}O}$, of the clumps derived from the 
\CIII\ data. By means of the
angular diameters in Col.~2 we have corrected $T_{\rm MB}$ for the beam 
filling factor, thus obtaining
the source-averaged brightness temperatures for the \CIII\ (1$-$0) and (2$-$1)
lines. The rotation temperature and \CIII\ total column 
density were computed from the line ratios assuming LTE conditions and 
optically thin lines (see e.g. Hofner et al.~\cite{hofner}).
 
\begin{figure}
\centerline{\includegraphics[angle=0,width=9cm]{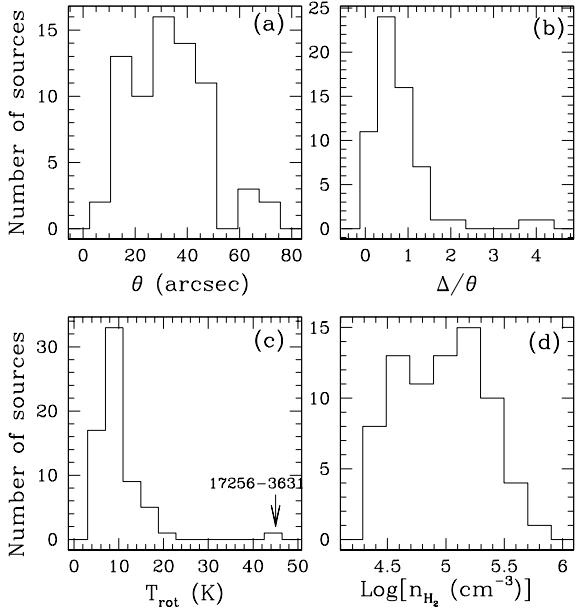}}
\caption{Histograms of some distance-independent parameters. 
(a) angular diameters
based on the SIMBA 1.2~mm continuum maps (HPBW$\simeq 21$\asec ). 
(b) ratio between angular separation between
the IRAS position and the position of the millimeter peak ($\Delta$) and 
the source angular diameter ($\theta$). (c) rotation temperatures
derived from \CIII\ line ratios. (d) H$_{2}$ volume densities derived 
from CS line ratios.}
\label{histo_t_n}
\end{figure}

We find rotation temperatures distributed around $\sim8-10$ K, with
the exception of 17256$-$3631, for which we obtain $T_{\rm rot}\sim 45$ K 
(see Fig.~\ref{histo_t_n} (c)). We have also computed the 
H$_{2}$ total column densities from $N_{\rm C^{17}O}$, assuming
a mean \CIII\ abundance relative to H$_{2}$ 
$X_{\rm C^{17}O}=N_{\rm C^{17}O}/N_{\rm H_{2}}\sim 3.9\times10^{-8}$ 
(Wilson \& Rood~\cite{wer}). 
They are found to be in the range $\sim 10^{22}-10^{24}$cm$^{-2}$. 
The values of $N_{\rm H_{2}}$ are of the same order as those found by
Hofner et al.~(\cite{hofner}) in their sample of UC \HII\ regions,
whereas those of $T_{\rm rot}$ are more than $\sim 2$ times lower.
We will discuss this result in Sect~\ref{tkin_discu}.

\subsubsection{H$_{2}$ volume density from CS lines}
\label{lvg}

We have used the LVG code of Cesaroni et al.~(\cite{cesa91}), with
the collisional rates from Turner et al.~(\cite{turner}),
to derive H$_{2}$ volume densities, $n_{\rm H_{2}}$, from the CS lines. 
The LVG code computes the ratio between the brightness temperature, 
$T_{\rm B}$, of the CS (2$-$1) and (3$-$2) transitions, as a function of 
the kinetic temperature, $T_{\rm k}$, the
H$_{2}$ volume density, the CS average abundance, and the velocity
gradient, ${\rm d}v/{\rm d}r$. By comparing our data with the prediction of 
the models, and assuming that $T_{\rm kin}\sim T_{\rm rot}$ from
\CIII , we can thus
derive an estimate of the H$_{2}$ volume density.

We have first calculated the
brightness temperature, $T_{\rm b}$, of the lines from the 
measured $T_{\rm MB}$ according to
the relation $T_{\rm b}=T_{\rm MB}(1+(\theta_{\rm MB}/\theta)^{2})$ (where
$\theta_{\rm MB}$ is the Telescope HPBW and $\theta$ is the source size),
and computed the corresponding line ratio.
Then, assuming a mean CS abundance of $10^{-8}$ (Irvine et al.~\cite{irvine}),
a velocity gradient of 10 \kms\ parsec$^{-1}$ (which is the
mean value of the ratio between the linewidths and the clump diameters), 
and the temperature obtained from
\CIII\ (Sect.~\ref{tkin}), we used 
the LVG code to estimate the value of $n_{\rm H_{2}}$
that could reproduce the ratio 
$T_{\rm B}{\rm [CS(2-1)}/T_{\rm B}{\rm [CS(3-2)]}$.
The same computation was repeated using
the dust temperature, $T_{\rm d}$, which will be derived in 
Sect.~\ref{tdust}.
For sources for which we could not derive the \CIII\ temperature we have used 
a representative temperature $T_{\rm rot}=10$ K (see Fig.~3c).
The values listed in Table~\ref{tdist_ind} are the geometric mean of these
two density estimates. The ratio between the two estimates is on average a 
factor of 3.
We obtain $n_{\rm H_{2}}\sim 10^{4.5}-10^{5.5}$ \cmc , as shown in
Fig.~\ref{histo_t_n} (d).

\subsection{Kinematic distances and distance-dependent parameters}
\label{dist}

Kinematic distances, $d$, are listed in Col.~2 of Table~\ref{tdist_dep}. They have been
estimated from the CS line velocity using the rotation curve of
Brand \& Blitz (\cite{b&b}). The method is
valid for distances from the galactic center between 2 and 25 kpc.
We could not assign any distance to four sources (15371$-$5458, 17230$-$3531, 
17410$-$3019 and 
17425$-$3017), because the corresponding distance estimates were out of this 
interval.

For sources inside the solar circle, two solutions for the kinematic distance
({\it near} and {\it far}) are possible. In a few cases, this ambiguity can 
be solved: for sources that would be more than 150 pc from the galactic
plane (i.e. twice the scale height of the molecular disk), the near 
distance was adopted. For one source (17040$-$3959),
the near distance implies a value of the dust mass (see Sect.~\ref{mass})
which is very unlikely ($\sim 0.02 M_{\odot}$), and hence the far distance 
was assumed, even though it is at 80 pc from the galactic plane. 
For all other sources we 
could not solve the distance ambiguity, and hence in Table~\ref{tdist_dep} 
we give both values.

The physical parameters which depend on the source distance are
listed in two tables: Table~\ref{tdist_dep} gives the clump linear
diameters, luminosities and dust temperatures (Col.~3, 4 and 5, respectively);
Table~\ref{tmass} lists
the masses estimated from dust emission (Col.~2), the 
virial masses (Col.~3), and the masses derived from the
\CIII\ and CS emission (Cols.~4 and 5, respectively). For sources with 
distance ambiguity, near and far estimates are listed in each 
column of both tables. 

\subsubsection{Linear diameters and luminosities}
\label{dist_lum}

\begin{figure}
\centerline{\includegraphics[angle=0,width=9cm]{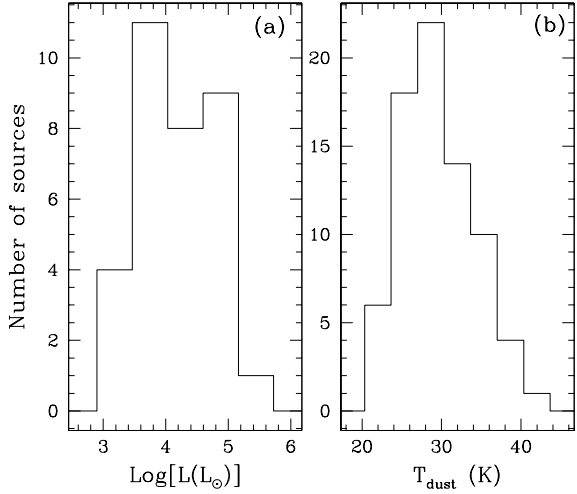}}
\caption{(a): histogram of the luminosities for sources
without distance ambiguity.
(b): histogram of the dust temperature of the clumps, derived from 
grey-body fits to the SEDs.}
\label{histo_l_td}
\end{figure}

The linear sizes have been computed from 
the angular diameters listed in Table~\ref{tdist_ind}, and are between
$\sim 0.1$ and $\sim 2$ pc, typical of clumps hosting
young high-mass objects (see e.g. Kurtz et al. \cite{kurtz}).
The luminosities were calculated by integrating the IRAS flux densities. The
contribution from longer wavelengths was taken into account by extrapolating
according to a black-body function that peaks at 100 $\mu$m and has the same
flux as the source at that wavelength. 
The distribution of the luminosities for sources without distance
ambiguity is shown in Fig.~\ref{histo_l_td} (a): we find
luminosities in the range $\sim 10^{3}-10^{5}L_{\odot}$.
For sources with distance ambiguity, in several cases
the {\it far} estimate is also of order $10^{6}L_{\odot}$. 
This confirms that the
embedded sources are indeed intermediate- or high-mass objects.

\subsubsection{Dust temperatures}
\label{tdust}

By fitting grey-bodies to the 1.2~mm continuum flux densities and 
the 60 and 100$\mu$m IRAS Point Source Catalog data, we have derived 
best-fit dust temperatures, $T_{\rm d}$. 

We find values of $T_{\rm d}$ distributed around $\sim 30$ K (see 
Fig.~\ref{histo_l_td} (b)), significantly 
higher ($\sim$ a factor 3$-$4) than the rotation temperatures $T_{\rm rot}$ 
estimated from \CIII\ lines (Sect.~\ref{tkin}). This difference is likely 
due to the fact that IRAS detects the 
emission of warm dust inside the innermost part of the clumps, whereas the 
\CIII\ (1$-$0) and (2$-$1) lines trace the more extended and colder envelope,
because of the low excitation of the $J=0,1,2$ levels ($\leq 16$ K).
\begin{figure}
\centerline{\includegraphics[angle=0,width=7cm]{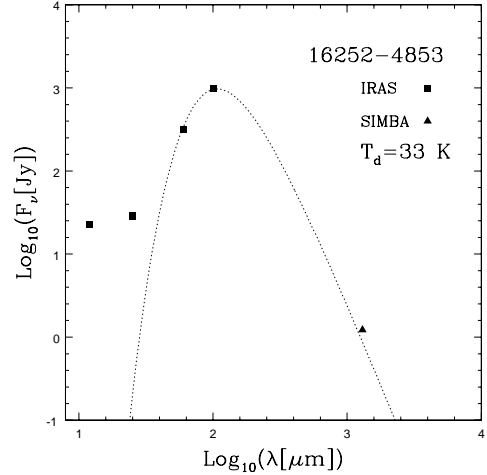}}
\caption{SED of 16252$-$4853. Symbols have the meaning indicated in 
the top
right-hand corner. The dotted line represents the best grey-body fit 
to points with $\lambda\geq 60\mu$m,
obtained for dust temperature of 33 K and dust opacity index $\beta=2$.}
\label{graybody}
\end{figure}

As previously said,
the values of $T_{\rm d}$ have been obtained by fitting only
the millimeter point and the 60 and 100 $\mu$m IRAS points of the 
SED (an example of a grey-body fit is shown in 
Fig.~\ref{graybody}). 
In fact, in almost all observed sources the SED shows a shape that 
cannot be fitted with a single grey-body, but rather two
grey-bodies: a ``cold'' one which fits the mm data and the IRAS 60 and
100 $\mu$m points, and a ``hot'' one, which fits the 12 and 25 $\mu$m data.
Various authors (see e.g. Sridharan et 
al.~\cite{sridharan}) have indeed
shown that flux densities measured in different bands of the IRAS 
Catalogue do not necessarily arise from the same region. 
Furthermore, Molinari et al.~(\cite{mol98b}) and
Fontani et al.~(\cite{fonta1}, \cite{fonta2}) have recently
demonstrated that in three {\it Low} sources of the sample 
selected by Palla et al.~(\cite{palla})
the 60 and 100 $\mu$m emission arises from a compact core likely
hosting a massive protostar, while the emission at 12 and 25 $\mu$m 
is due to a cluster of more evolved IR sources surrounding the core.
Therefore, the values of $T_{\rm d}$ listed in Table~\ref{tdist_dep},
derived by fitting only the points with $\lambda\geq60\mu$m, are 
representative of the cold dust component.

In the fits we have assumed a dust opacity
$\kappa_{\nu}=\kappa_{\rm 230 GHz}(\nu ({\rm GHz})/230)^\beta$, where 
$\kappa_{\rm 230 GHz}=
0.005\;{\rm cm^{2}g^{-1}}$, which implies
a gas-to-dust ratio of 100 (Kramer et al.~\cite{kramer}). We have also 
assumed $\beta=2$
which is a typical value derived for dusty envelopes of
massive (proto)stellar objects (Hunter~\cite{hunter}; Molinari et 
al.~\cite{mol00}).

We stress that this is a simplified approach,
since these regions can be very complex (as demonstrated by
various authors, see e.g. Fontani et al.~\cite{fonta1},~\cite{fonta2}) 
and a proper modeling of the SED would require many more details
(source geometry, contribution of very small dust grains and PAHs).
However, this approach would require a substantial number of assumptions 
and would go beyond the scope of this paper. 

Our values of $T_{\rm d}$ are similar to those of the {\it Low} sources
studied by Molinari et al.~(\cite{mol00}), who derived in the 
same way the temperature of the cold dust in their sources. A similar analysis
has been made by
Sridharan et al.~(\cite{sridharan}), but they fitted the SED with two 
grey-bodies. We stress that, even if our estimates of 
$T_{\rm d}$ have been derived from a single grey-body, 
fitting the SED with two grey-bodies does not significantly affect
the parameters that one derives from a single grey-body fit, because
the two components refer to well-separated parts of the spectrum: 
typically, the correction would be $<5\%$. Hence, our dust temperatures 
can be compared to those obtained by Sridharan et al.~(\cite{sridharan}):
they derive an average value for $T_{\rm d}$ of the ``cold'' grey-body
of $\sim 50$ K, which is
$\sim 1.6$ times larger than that derived in this work for our sources.
\subsubsection{Mass estimates}
\label{mass}

Clump masses have been estimated using 4 different methods:
\begin{itemize}
\item One can compute the total mass from the dust millimeter-continuum
emission ($M_{\rm cont}$) assuming optically thin conditions and
a constant gas-to-dust ratio.
Clump masses from mm continuum have also been derived by Beuther et 
al.~(\cite{beuther}) for the {\it High} sources of the
Sridharan/Beuther sample (see Sect.~\ref{intro}). 
In order to make a consistent comparison with the masses derived
by those authors, we compute $M_{\rm cont}$ with the relation
used by them (see their Sect.~3.2):
\begin{eqnarray}
M_{\rm cont}(M_{\odot})&=&1.3\times 10^{-3}F_{\nu}({\rm Jy})[d({\rm kpc})]^{2}\nonumber \\  
& & \left[{\rm exp}\left(\frac{h\nu}{kT_{\rm d}}\right)-1\right]\left(\frac{\nu}{\rm 2.4 THz}\right)^{-3-\beta}
\label{emdust}
\end{eqnarray}
This relation has been derived adopting the same dust opacity as used
in Sect.~\ref{tdust}, which assumes a gas-to-dust ratio of 100. We have 
also used the same opacity index, $\beta=2$.
\item From the line FWHM and the clump linear diameters, we can estimate 
the virial mass, $M_{\rm vir}$. Assuming that the source is spherical and 
homogeneous, and neglecting the contributions of magnetic field and surface 
pressure, the virial mass is given by:
\begin{equation}
M_{\rm vir}(M_\odot)=0.509\,d({\rm kpc})\,\theta({\rm arcsec})\,\Delta v^{2}({\rm km/s})    \label{emvir}
\end{equation}
where $d$ is the source distance (Table~\ref{tdist_dep}), $\theta$ the 
source angular diameter (Table~\ref{tdist_ind})
and $\Delta v$ is the FWHM of the CS (2$-$1) line (Table~\ref{tcsline}). 
\item From the H$_{2}$ total column density derived from the \CIII\ lines 
(Sect.~\ref{tkin}) we may deduce the gas mass according to the expression:
\begin{equation}
M_{\rm C^{17}O}=\frac{\pi}{4}{D^{2} N_{\rm H_{2}}m_{\rm H_{2}}}     
\label{emcd}
\end{equation}
where $D$ is the source diameter in Table~\ref{tdist_dep}, $m_{\rm H_2}$ the 
mass of the H$_2$ molecule, and $N_{\rm H_{2}}$ is the beam-averaged column 
density (Table~\ref{tdist_ind}).
\item Assuming a spherical source, we can compute the gas mass from the 
CS lines, $M_{\rm CS}$, using the H$_{2}$ average
volume density derived from LVG calculations (see Sect.~\ref{lvg})
and the source diameters given in Table~\ref{tdist_dep}.
\end{itemize}
All mass estimates are listed in Table~\ref{tmass}.
We find clump masses ranging from a few tens $M_{\odot}$
up to $\sim 10^{5}M_{\odot}$.
We will discuss the different mass estimates in 
Sect.~\ref{stability}.

In Sect.~\ref{contdata} we pointed out that a few sources have been
detected in \CIII\ but not in the millimeter continuum, and that this
can be due to a ``distance effect''. We can now justify this
statement. From Eq.~(\ref{emdust}) one
can estimate the continuum flux expected for a clump with mass $M$  
located at the far distance, and compare this with the sensitivity
of our maps, to check if the emission is not detected because
the clump is too far away. Since we have no direct estimate
for the clump masses of these sources, we have assumed a
representative value 
of $10^{3}$ $M_{\odot}$: we infer that, at the far distances given in 
Table~\ref{tdist_dep}, for 16153$-$5016, 16254$-$4844 and 16417$-$4445 the 
expected
fluxes are $\sim 0.07$ Jy, while for 15038$-$5828 and 16403$-$4614 they are 
$\sim 0.15$ and $\sim 0.13$ Jy, respectively. These values are comparable
to the 3$\sigma$ level in the maps, which is 
$\geq 0.07$ Jy beam$^{-1}$. We hence conclude that the non detection
of these sources in the continuum could be due to our sensitivity
limit and the fact that they are located at the far distance.

\section{Discussion}
\label{discu}

The most important result of this work is that a large fraction ($\sim85\%$)
of the sample is associated with dense gas, as partially expected on the
basis of the criteria applied to select our sources.
In Sect.~\ref{intro} we have stressed that
this work is the extension to $\delta<-30^{\circ}$ of the project 
started by Palla et al.~(\cite{palla}) in the northern hemisphere, aimed at 
the identification
of precursors of UC \HII\ regions through a comparative study of 
{\it High} and {\it Low} sources.
For this reason, in the following we will compare the properties 
derived in Sects.~\ref{res} and \ref{phipar} to those of
other well known samples of high-mass protostellar candidates and 
massive YSOs with IRAS colours typical of {\it High} sources. 

\subsection{Rotation temperatures from \CIII\ }
\label{tkin_discu}

The rotation temperatures derived in Sect.~\ref{tkin} from \CIII\
are distributed around $\sim 8-10$ K.
Molinari et al.~(\cite{mol96}) found a mean value 
of $\sim 22$ K for their sources, without any 
significant difference between {\it High} and {\it Low} sources, and 
Sridharan et al.~(\cite{sridharan})
found an average value of 19 K in their sample of massive protostar  
candidates. Furthermore, Brand et al.~(\cite{brand}) measured
the rotation temperature in clumps associated with 6 {\it Low} sources, 
finding temperatures from $\sim 20$ K to $\sim 50$ K.
However, both Molinari et al.~(\cite{mol96}) and 
Sridharan et al.~(\cite{sridharan}) used NH$_{3}$ lines in deriving 
$T_{\rm rot}$, while Brand et al.~(\cite{mol96}) used \ace\ lines,
which probably trace a different region of the clumps. 

Hofner et al. (\cite{hofner}) made
a survey of \CIII\ towards UC \HII\ regions. They observed the 
\CIII\ (1$-$0), (2$-$1) and (3$-$2) lines with the IRAM$-$30m 
and the KOSMA$-$3m
telescopes. The authors found temperatures from 13 to 41 K, with a
mean value of $\sim23$ K. This value is higher than that derived by us. 
However, it must be noted that the \CIII\ (3$-$2) 
transition likely arises from a more internal and hotter region than
that traced by the (1$-$0) and (2$-$1) lines.
To allow a consistent comparison we 
have derived the rotation temperature of the sources of the Hofner et al. 
(\cite{hofner}) sample using only the
transitions observed by us, namely the \CIII\ (1$-$0) and (2$-$1) lines:
we thus obtain temperatures of $\sim 20$ K on average. This value
is marginally lower than that obtained by Hofner et al.~(\cite{hofner}), 
and still $\sim 2$ times higher than ours. 

A possible explanation of this is that our sources are on average 
less luminous than the UC \HII\ regions observed by Hofner et 
al.~(\cite{hofner}). In fact, in clumps
where the gas is heated by an embedded (proto)star with luminosity $L$, 
the gas temperature
at a distance $r$ from the central (proto)star is
expected to scale as (see e.g. Doty \& Leung~\cite{doty})
$T\propto (L^{1/2}/r)^{\alpha}$, where ${\alpha}$ typically
varies between 0.3 and 0.5. Assuming coupling between gas 
and dust, which is plausible for densities of $\sim10^{5}$\cmc , this
holds also for the dust temperature. Therefore, sources with 
higher luminosities are expected to be hotter at the same distance from 
the central object.
Another possible explanation could be the role of the different
angular resolution of the observations: Hofner et 
al.~(\cite{hofner}) observed the \CIII\ (1--0) and (2--1) lines with
an angular resolution two times better than ours. Therefore, they 
were observing lines arising from a more internal, and probably hotter,
region of the clumps. On the basis of our data, it is impossible
to discriminate between the two hypothesis presented above.

\subsection{Linewidths}
\label{linewidths}

\subsubsection{Comparison with UC \HII\ regions}
\label{line-uchii}
 
\begin{figure}
\centerline{\includegraphics[angle=0,width=8.5cm]{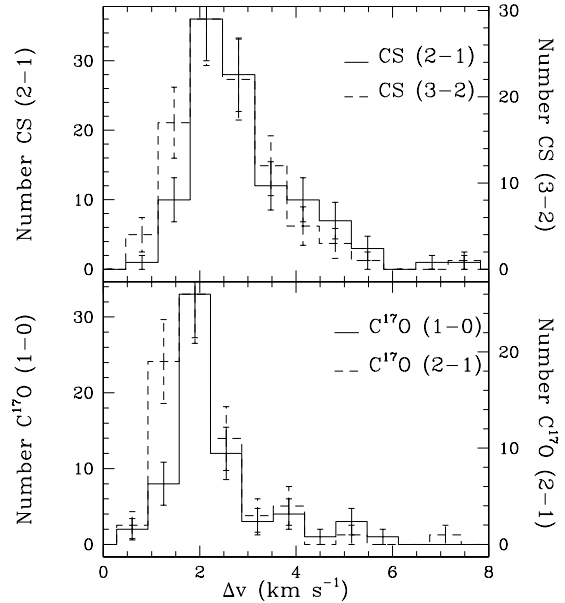}}
\caption[Distribution of the CS linewidths]{Top panel:
Histogram of the CS linewidths. Bottom panel: same as top panel for the \CIII\ linewidths.}
\label{histo_dv}
\end{figure}

In Fig.~\ref{histo_dv} we show the 
distributions of the linewidths ($\Delta v$) of the CS and \CIII\ 
lines (top and bottom panels, respectively), measured in 
Sects.~\ref{csdata} and \ref{c17odata}.
For the CS lines $\Delta v$ is on average $\sim 2.7$ \kms , with no
significant difference between the (2$-$1) and (3$-$2) transitions.
Such a mean value is much lower than that measured by 
Cesaroni et al.~(\cite{cesa91}) towards a sample of UC \HII\ regions: they
found linewidths from $\sim 3.5$ \kms\ to $\sim 9$ \kms , with an
average value of $\sim 6$ \kms\ in both 
transitions. For the \CIII\ lines, the mean value observed by us is 
$\Delta v \sim 2$ \kms\ (see 
bottom panel of Fig.~\ref{histo_dv}), $\sim 3$ times lower than that found 
by Hofner et al. (\cite{hofner}), from observations of UC \HII\ regions.

A possible interpretation of these results is that the turbulence is much
lower in our clumps than in those associated with UC \HII\ regions. 
The turbulence in high-mass star forming regions is due to a 
variety of phenomena (e.g. powerful outflows, winds, infall), and 
is correlated to the activity of the embedded objects: 
less evolved objects are thought to be associated with more
quiescent envelopes. Therefore, the narrower lines found in our sources
suggest that the embedded objects do not contain
already formed stars. This interpretation is also supported by the results
of Brand et al.~(\cite{brand}), who came to the same conclusion
for their sample of northern {\it Low} sources.

\subsubsection{Comparison with {\it High} sources}
\label{line$-$high}

\begin{figure}
\centerline{\includegraphics[angle=0,width=8.5cm]{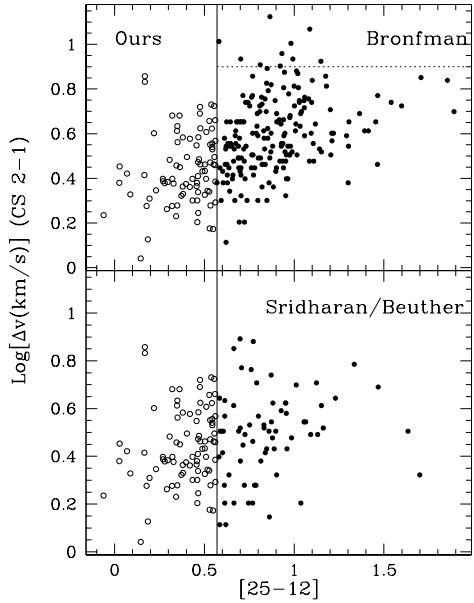}}
\caption[$\Delta v$ of the CS (2$-$1) transition against the
(25$-$12) color index]{Plot of the linewidths of the CS (2$-$1) 
transition against the
[25$-$12] color index. Top panel: open circles represent {\it Low} 
sources observed by us; filled circles indicate potential UC \HII\
regions observed by Bronfman et al.~(\cite{bronfman}), 
observed with the SEST, with $\delta<-30^{\circ}$ and 
satisfying the criteria adopted by Palla et al.~(\cite{palla}): 
hence they are {\it High} sources. The dashed line indicates the 
maximum value of $\Delta v$ measured
for the sources of the Sridharan/Beuther sample. 
Bottom panel: same as Top panel for the sources 
observed by Sridharan et al.~(\cite{sridharan}).}
\label{dv_color}
\end{figure}

In the top panel of Fig.~\ref{dv_color} we plot the
linewidth of the CS (2$-$1) line against the [25$-$12] colour, and compare 
our data with those of Bronfman et al~(\cite{bronfman}).
In order to make a consistent comparison, we have plotted only 
sources that have $\delta<-30^{\circ}$ (namely
sources which have all been observed with the SEST), 
and that satisfy
the criteria adopted by Palla et al.~(\cite{palla}) to identify compact
molecular clouds: since all Bronfman sources have colour indices 
[25$-$12]$\geq0.57$ and [60$-$12]$\geq1.3$ (see Sect.~\ref{intro}), 
the subsample selected by
us consists of {\it High} sources. Hereafter, this subsample (190 sources)
will be called ``Bronfman sample''.
It is worth noting that we cannot take out the \HII\ regions from this
sample because of the lack of extensive surveys of \HII\ regions south
of $\delta=-30^{\circ}$.
The mean value of the data of the Bronfman sample is $\sim 3.9$ \kms 
(median=3.9 \kms ), 
with a standard deviation $\sigma\simeq 1.5$ \kms , while for our sample
we find a mean value of $\sim 2.7$ \kms (median=2.6 \kms ), with a 
standard deviation $\sigma\sim 1.4$ \kms .

In the bottom panel, we show a comparison between our sample and the
69 {\it High} sources of the ``Sridharan/Beuther sample'' (see 
Sect.~\ref{intro}). 
The mean value that we derive for the Sridharan/Beuther sample
is $\sim 3.1$ \kms\ ($\sigma\simeq 1.5$ \kms ), and the median is 3.2 \kms .
From a purely statistical point of view, the linewidth distributions
of the three samples are mutually consistent among them.
However, from Fig.~\ref{dv_color} one can notice that our 
sources and those of the Sridharan/Beuther 
sample have similar linewidths, whereas a small fraction 
(11 out of 190 sources)
of the {\it High} sources of the Bronfman sample have linewidths larger
than those measured by us and by Beuther et al.~(\cite{beuther}). 
More quantitatively, we find that approximately $10\%$ of the sources in the
Bronfman sample have $\Delta v >7$ \kms , 
while these percentages are $< 5$ and $< 1$
in the Sridharan/Beuther and in our sample, respectively.
This may suggest that the {\it High} sources of the Bronfman sample 
have CS linewidths slightly different from
those of the Sridharan/Beuther sample. However, as previously pointed out,
the Bronfman sample also contains \HII\ regions, which have 
been excluded from the Sridharan/Beuther sample.
For this reason, we believe that the sources 
of the Bronfman sample which show larger linewidths are likely associated 
with \HII\ regions. We conclude that {\it High} and {\it Low} sources 
not associated with \HII\ 
regions have similar linewidths.

\subsubsection{Full width at zero intensity of the lines}

The full width at ``zero intensity'' (FWZI) of a line 
provides information
about the presence of non-Gaussian wings, and hence of an outflow.
Since bipolar outflows are believed to be strictly related to the
accretion process of forming stars of both low- and high-mass (see e.g. 
Beuther et al.~\cite{beuther2}), it is important to check the
association of our sources with an outflow to better understand
their nature.
For this reason, we have measured the FWZI of the CS lines and
compared them to the theoretical values expected from purely Gaussian 
lines. In the ideal case of a spectrum without noise,
the wings of a Gaussian line asymptotically approach zero,
and the FWZI tends to infinity.
In the real spectra, the ``zero intensity'' depends on the noise 
level, and therefore the measured FWZI depends both on the FWHM 
and on the signal-to-noise ratio $T_{\rm max}/\sigma$, where 
$T_{\rm max}$ is the line peak and $\sigma$ is the rms noise in the
spectrum.
Taking as ``zero intensity'' the 3$\sigma$ level, 
one can demonstrate that 
the FWZI is related to the FWHM and the signal-to-noise ratio as follows:
\begin{equation}
\rm FWZI_{\rm gauss}=\frac{FWHM}{\sqrt{\rm ln2}}\sqrt{\rm ln\left[\frac{T_{\rm max}}{3\sigma}\right]}\:.
\label{efwzi}
\end{equation}

Deviations from this relationship are due to 
non-Gaussian wings, and may hence indicate the presence of an outflow. 

\begin{figure}
\centerline{\includegraphics[angle=0,width=9.5cm]{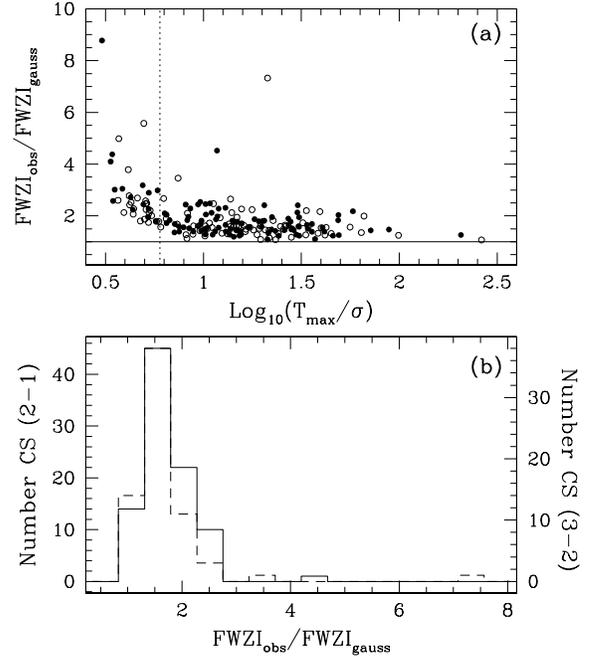}}
\caption{(a) ratio between the observed full width at zero intensity 
(FWZI$_{\rm obs}$) and that expected for a line with Gaussian shape
(FWZI$_{\rm gauss}$) from Eq.~(\ref{efwzi}) for the 
CS (2$-$1) (filled circles) and (3$-$2) (empty circles) lines 
versus the signal-to-noise ratio ($T_{\rm max}/\sigma$) of the spectra. 
Significant 
data are those with $T_{\rm max}/\sigma>6$ 
(to the right of the dotted line).
(b) histograms of the quantity $FWZI_{\rm obs}/FWZI_{\rm gauss}$
for both CS (2--1) (solid line) and (3--2) (dashed line) lines. 
Only sources with $T_{\rm max}/\sigma>6$ are considered here.}
\label{fwzi_sn}
\end{figure}

In Fig.~\ref{fwzi_sn} (a) we plot the ratio between the observed FWZI
(FWZI$_{\rm obs}$) and that expected from Eq.~(\ref{efwzi}) as a function 
of the signal-to-noise ratio, both for the 
CS (2$-$1) and (3$-$2) lines.
The values of FWZI$_{\rm gauss}$ have been computed from
the FWHM listed in Table~\ref{tcsline}.
The FWZI$_{\rm obs}$ are computed as the separation between the first 
channels on the right and 
left from the line peak with intensity lower than 3$\sigma$. 
The significant data are those with line intensity at half 
maximum $>3\sigma$, i.e. with $T_{\rm max}/\sigma$$>6$. 
As expected, all observed
lines have FWZI$_{\rm obs}$/FWZI$_{\rm gauss}\geq 1$.
Fig.~\ref{fwzi_sn} (b) shows that the majority ($\sim 70\%$) 
of our lines 
with $T_{\rm max}/\sigma>6$ has 
FWZI$_{\rm obs}\geq 1.5$ FWZI$_{\rm gauss}$,
suggesting the presence of outflows in many of our sources.
The ``most frequently occurring value'' of FWZI$_{\rm obs}$ is 
$\sim 6$ \kms\ for both
the CS (2$-$1) and (3$-$2) lines, in good agreement with the values
found by Brand et al.~(\cite{brand}) in a sample of 11 
northern {\it Low} sources, in which they found
an average value of 5.9 \kms\ for the CS (3$-$2) line.

\subsection{Luminosities}
\label{lum}

\subsubsection{Comparison with {\it High} sources}
\begin{figure}
\centerline{\includegraphics[angle=0,width=9cm]{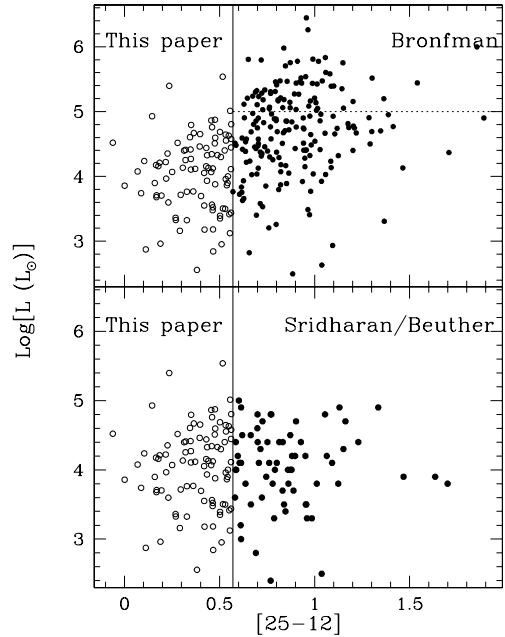}}
\caption[Bolometric luminosity against the
(25$-$12) color index]{Plot of the bolometric luminosity against the
[25$-$12] color index. For sources with distance ambiguity the
near value has been adopted. Top panel: open circles represent 
{\it Low} sources observed by
us; filled circles indicate {\it High} sources observed by Bronfman et 
al.~(\cite{bronfman}), with $\delta<-30^{\circ}$. The dotted line 
indicates the maximum value of $L_{\rm near}$ found in the Sridharan/Beuther
sample.
Bottom panel: same as top panel for the sources of the Sridharan/Beuther
sample.}
\label{lum_tot}
\end{figure}

In order to see if {\it High} and {\it Low} sources are associated with 
young stars of different masses, we verify if there is 
any significant difference in
luminosity between our sources and those of the other samples
of high-mass protostellar candidates.
In Fig.~\ref{lum_tot} we compare
the bolometric luminosities of our sources with those of the Bronfman sample
(top panel), selected as explained in Sect.~\ref{line$-$high} and 
those of the Sridharan/Beuther sample (bottom panel). One can see that 
the sources in our sample and in that of Sridharan/Beuther have similar 
luminosities, mostly distributed between $10^{3}$ and 
$10^{5}L_{\odot}$ (only two
of them have $L>10^{5}L_{\odot}$).
This is consistent with the results of Palla et 
al.~(\cite{palla}), who noted that the luminosity distributions
for the sources of the {\it High} and {\it Low} groups with 
$\delta\geq-30^{\circ}$ are similar. On the other hand, 
in the Bronfman sample, which also includes \HII\ regions, 
$\sim 30\%$ of the sources have luminosities larger than $10^{5}L_{\odot}$.
Therefore, the sources with the highest bolometric
luminosities are likely associated with \HII\ regions (or equivalently
with more massive stars), 
which are expected to be brighter at FIR wavelengths.

\begin{figure}
\centerline{\includegraphics[angle=0,width=9.3cm]{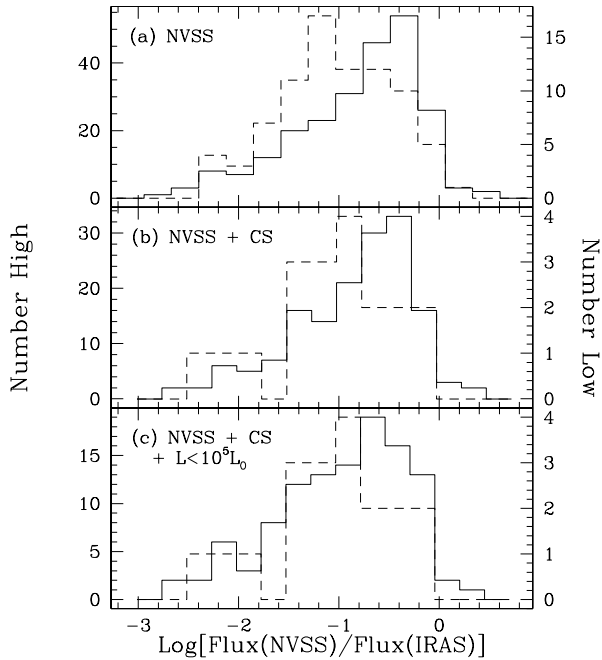}}
\caption{Distribution of the ratio between
the NVSS radio flux and the IRAS integrated flux for {\it High} (solid
line) and {\it Low} (dashed line) sources. (a) Sources detected in NVSS;
(b) Same as panel (a) for sources detected also in dense gas 
(CS for {\it Highs}, CS or \AMM\ for {\it Lows}; (c) Same as panel 
(b) for sources with luminosities lower than $10^{5}L_{\odot}$.}
\label{fnvss}
\end{figure}

This explanation is further supported by the plots shown 
in Fig.~\ref{fnvss}, in which
we present the distribution of the ratio between
the radio-continuum flux,
taken from the on-line NRAO VLA Sky Survey (NVSS) database
\footnote{The NVSS data are available at http://www.cv.nrao.edu/nvss/ }, 
and the IRAS integrated flux for both {\it Highs} 
and {\it Lows} detected in the NVSS. The NVSS surveyed the sky
north of $\delta=-40^{\circ}$ at 1.4~GHz, with an angular
resolution of $\sim 45$\asec . For further details about the NVSS data,
see Condon et al.~(\cite{condon}). 

In Fig.~\ref{fnvss} (a) we have plotted
all sources which satisfy the colour-colour criteria by Palla et 
al.~(\cite{palla}) belonging both to the northern and the
southern sky. 
In Fig.~\ref{fnvss} (b) we have plotted only those also associated 
with dense gas, i.e. detected
in CS by Bronfman et al.~(\cite{bronfman}) ({\it Highs}), and detected in 
CS in this work ({\it Lows}).
Since the number of {\it Lows} detected by us and in the NVSS
was low (only 6), 
we have included in the analysis the {\it Low} sources
detected in \AMM\ by Molinari et al.~(\cite{mol96}).
The mean values of the NVSS-to-IRAS flux ratios are
$\sim 0.3$ (with standard deviation $\sigma\sim 0.3$) for {\it High} sources,
and $\sim 0.2$ ($\sigma\sim 0.2$) for {\it Low} sources,
for the distributions of both Fig.~\ref{fnvss} (a) and (b).
Hence, {\it Highs} and {\it Lows} have similar distributions
of the NVSS-to-IRAS flux ratios.
Although these values are very similar, the offset between the
peaks of the distributions of {\it Highs} and {\it Lows} in 
Figs.~\ref{fnvss} (a) and (b) leds us to speculate that a fraction of  
the {\it High} sources has higher NVSS-to-IRAS flux ratios than 
the {\it Low} sources, suggesting that the former group might be 
more tightly associated with \HII\ regions.
It is also worth noting that the distributions shown in 
Figs.~\ref{fnvss} (a) and (b) contain sources of {\it all} 
luminosities. In Fig.~\ref{fnvss} (c) 
we have plotted only sources detected in dense gas 
with $L<10^{5}L_{\odot}$: this allows us to make a consistent comparison
given the lack of sources with $L>10^{5}L_{\odot}$ in the {\it Low}
sample. Although the uncertainties are very large
because the statistics are poor, especially for the {\it Low} sources,
the NVSS-to-IRAS flux ratios are distributed similarly for the two
groups: the mean values are $\sim 0.24$ and $\sim 0.19$ 
($\sigma\sim0.2$) for {\it Highs} and {\it Lows}, respectively.
This means that the most luminous sources of the {\it High} group 
(which are excluded from this diagram)
have the highest NVSS-to-IRAS fluxes, and thus are likely  
associated with evolved \HII\ regions.

For this reason, we believe that the luminosities of {\it Low} and 
{\it High} sources {\it not associated} with \HII\ regions are similar, 
and that the embedded high-mass objects likely have similar
mass. This conclusion supports
the results of the previous studies made by Molinari
et al.~(\cite{mol98a},~\cite{mol00}) of the {\it Highs} and {\it Lows}
of the northern hemisphere.

\subsubsection{Mass-luminosity ratio and age of the sources}

\begin{figure}
\centerline{\includegraphics[angle=0,width=8.5cm]{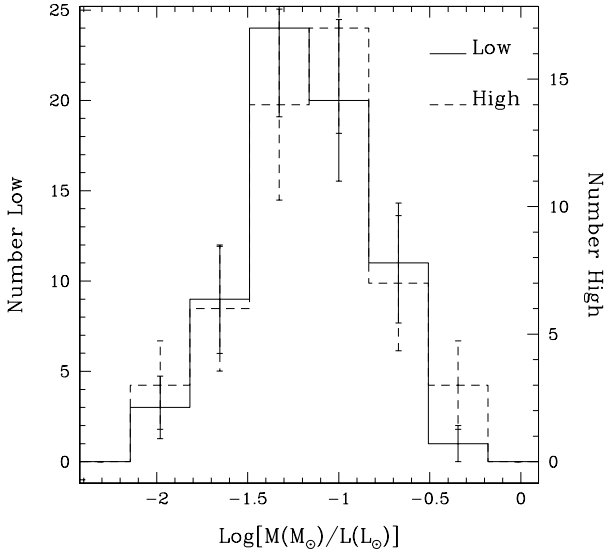}}
\caption[Distribution of $M/L$ for our sources and the {\it High} of the 
Sridharan sample]{Distribution of the distance-independent quantity 
$M/L$, where $M$ is the clump mass derived from dust emission, for
our {\it Low} sources (solid line) and the {\it High} sources
of the Sridharan/Beuther sample (dashed line). For both 
distributions, the mean value is $\sim 0.07M_{\odot}/L_{\odot}$.}
\label{m_l}
\end{figure}

Another important parameter for establishing the age of a clump
is the ratio between the mass of the
clump and the corresponding luminosity, $M/L$. This is believed to decrease
with time because during the star formation process more and more gas is
converted into stars. Therefore, for clumps with {\it comparable masses},
the ratio $M/L$ is an estimate of the degree of evolution of the embedded
source.  
With this in mind, in Fig.~\ref{m_l}, using the clump mass derived from 
dust emission, $M_{\rm cont}$, we plot the histograms
of the distance-independent
quantity $M/L$ for our {\it Low} sample, and the {\it High} sample of 
Sridharan/Beuther. No significant
difference is seen between {\it High} and {\it Low} sources. 
Sridharan et al.~(\cite{sridharan}) have compared their
sources to known UC \HII\ regions, finding a lower $M/L$ ratio in the 
latter. The clump masses of both samples were comparable, so that
the authors interpreted this result as an indication of the relative youth
of the sources of their sample. Since
the clumps associated with our {\it Low} sources also have masses 
similar to those associated with the UC \HII\ regions analysed
by Sridharan et al.~(\cite{sridharan}) and the {\it Highs}
of the Sridharan/Beuther sample, the 
different $M/L$ ratio can be interpreted the same way, namely that
our sources, as well as those of the Sridharan/Beuther sample, are 
younger than UC \HII\ regions.

\begin{figure}
\centerline{\includegraphics[angle=0,width=8.5cm]{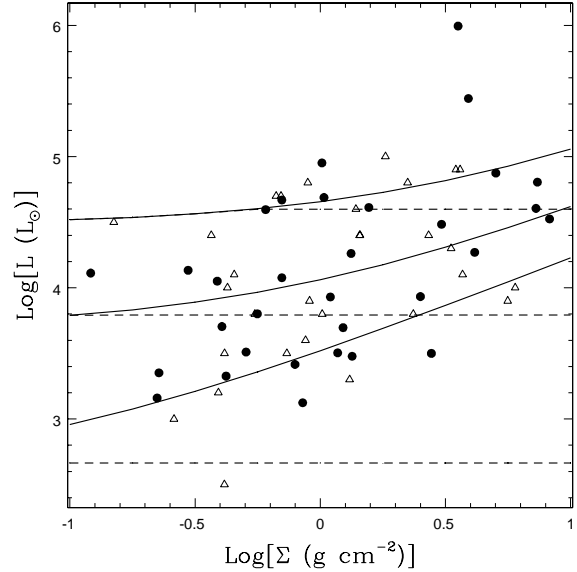}}
\caption{Plot of the bolometric luminosity, $L$, of our sources
and those of the Sridharan/Beuther sample
versus the gas surface density derived from dust emission,
$\Sigma=M_{\rm cont}/\pi R^{2}$. Filled circles
represent our {\it Low} sources; Open triangles correspond to
the {\it Highs} of the Sridharan/Beuther sample. The solid lines
are theoretical curves for an accreting protostar (from top to bottom)
with mass 20, 10 and 5 $M_{\sun}$, assuming the ZAMS radius
as protostellar radius. The dashed lines are the predictions 
obtained assuming the protostellar radius from Nakano et 
al.~(\cite{nakano}).}
\label{sigma_l}
\end{figure}

In accretion-dominated models of the evolution of a massive
protostar, one might expect to see correlation between the
properties of the core from which the protostar is accreting and
protostellar characteristics such as luminosity and outflow rate.
From the recent models of Tan~(\cite{tan}) (see also McKee \& Tan
~(\cite{mckeetan}),
one sees that there is a close connection between the accretion
rate onto a protostar and the column density of the clump in which
it forms (or eqivalently the surface density
$\Sigma = M_{\rm cont}/\pi R^2$).
Since the protostellar luminosity is partially due to
accretion, it seems reasonable to examine the dependence of
bolometric luminosity upon surface density for both our
sample and the Sridharan/Beuther sample.
The results (for sources without distance ambiguity)
are shown in Fig.~\ref{sigma_l} where we also show theoretical
predictions based on the results of Tan~(\cite{tan}) and Nakano et al.
(\cite{nakano}).
One sees that although there is a lot of scatter, there is
a tendency for an increase in protostellar luminosity with
clump column density. One also sees that in this diagram the {\it High}
and {\it Low} samples behave essentially in the same fashion.

It is interesting moreover that there is rough
agreement between the observations and predictions of the
models for assumed protostar masses in the range 5-20 $M_{\odot}$
where we have assumed half of the final protostellar mass to have
been accreted. We note also that there is
a variety of assumptions involved in deriving
the "theoretical curves" including the assumption that the luminosity
is dominated by the most massive protostar of what
presumably is an embedded cluster. Another uncertainty
involves the protostellar radius. To make this
point clear, we show
results in Fig.~\ref{sigma_l} for two extreme assumptions concerning
the protostellar radius: the value on the ZAMS and that derived 
according to the prescription
of Nakano et al.~(\cite{nakano}). We note that while the former appears to
give better agreement with the data of Fig.~\ref{sigma_l}, 
the latter is probably preferable both for theoretical reasons
discussed by Nakano et al. and
because of the fact that the predicted Lyman
continuum luminosities are much lower with this
assumption, consistent with the observation of little or no
centimeter continuum emission. In this case however, the 
predicted dependence
of luminosity upon $\Sigma $ is completely flat because the
protostellar radius becomes proportional to the
accretion rate and we need other indicators (such as outflow)
to test the hypothesis that we are observing accreting
protostars.

\subsection{Mass comparison and stability of the clumps}
\label{stability}

\begin{figure}
\centerline{\includegraphics[angle=0,width=8cm]{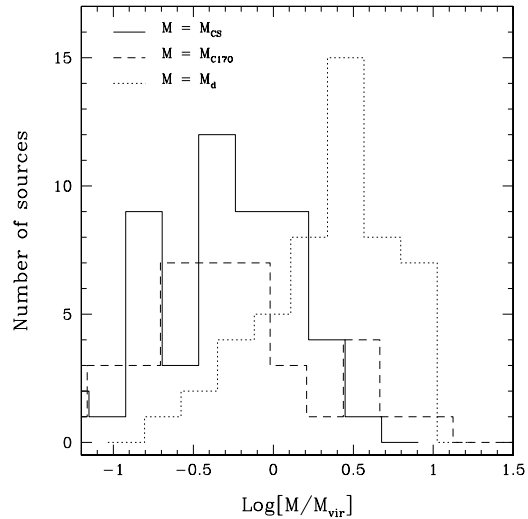}}
\caption[]{Histograms of the 
ratio between the virial mass $M_{\rm vir}$ and the clump masses
estimated with other methods: from CS ($M_{\rm CS}$, solid line), from 
\CIII\ ($M_{\rm C^{17}O}$, dashed line)
and from dust emission ($M_{\rm cont}$, dotted line).}
\label{isto_masse}
\end{figure}

The masses estimated from the 1.2~mm continuum, the CS and \CIII\ lines 
are compared to those deduced from the virial equilibrium 
(see Table~\ref{tmass}) in Figs.~\ref{isto_masse}, ($\sigma=0.3$),
in which we present
histograms of the ratio between the virial mass and the mass estimated
with the other methods. The average ratio between $M_{\rm CS}$ and 
$M_{\rm vir}$ is $\sim 0.8$ ($\sigma=0.7$),
and between $M_{\rm C^{17}O}$ and 
$M_{\rm vir}$ is $\sim 0.5$ ($\sigma=0.8$). The only mass 
estimate to be significantly different from the
others is the mass obtained from the 1.2~mm continuum, as 
demonstrated by Fig.~\ref{isto_masse}: the mean
ratio between $M_{\rm cont}$ and $M_{\rm vir}$ is $\sim 3.3$, with
a standard deviation $\sigma\sim 2.7$.

However, it must be noted that the virial mass, the CS mass and the \CIII\
mass have been obtained from
the physical parameters deduced from the {\it lines}
and the diameter of the {\it continuum} region:
they are hence ``hybrid'' quantities, and thus prone to 
unpredictable uncertainties.
Furthermore, $M_{\rm vir}$ was calculated for homogeneous clumps:
various authors (see e.g. Hatchell et al.~\cite{hatchell}, Beuther et al.
~\cite{beuther}, Fontani et al.~\cite{fonta02}) have shown that
clumps associated with high-mass YSOs have density
distributions described by a power-law of the type
$n\propto r^{-p}$, with $p$ typically ranging from 
$\sim 1.5$ to $\sim 2.5$. 
Such density profiles can significatively affect the
estimates of $M_{\rm vir}$ (see MacLaren~\cite{maclaren}): 
for example, for $p=2$, $M_{\rm vir}$ becomes
a factor $\sim 1.6$ smaller. Thus, the ratio $M/M_{\rm vir}$ can
also be affected by steep density profiles in the clumps.

A better estimate would require knowledge of the diameter of the line
emitting region, which is not available. 
However, one can compare our results to those obtained by Brand et 
al.~(\cite{brand}), who mapped 11 {\it Low} sources with 
$\delta\geq-30^{\circ}$ in various molecular lines, among which 
CS (3$-$2). From the CS (3$-$2)
lines they found a gas-to-virial mass ratio lower than one, 
consistent with our result for $M_{\rm CS}/M_{\rm vir}$ and
$M_{\rm C^{17}O}/M_{\rm vir}$. Therefore, we can reasonably
conclude that, within the uncertainties, our clumps could be
virialized.
\subsection{Where are the protostars?}

The results discussed in Sects.~\ref{linewidths} and \ref{lum}
show that in several respects (linewidth distribution, 
luminosity distribution, mass-luminosity ratio, NVSS-to-IRAS flux ratio) 
the {\it Low} and 
{\it High} sources with luminosity $L<10^{5}L_{\odot}$ are very similar.
In particular, both samples seem to be
associated with high-mass protostellar objects. 
One might be tempted to conclude therefore that the IRAS colours,
on which the distinction between {\it Highs} and {\it Lows}
is based, are irrelevant
for determining the evolutionary stage of these objects. 
At a first glance, this result seems to contradict the 
conclusions of previous studies of {\it Highs} and {\it Lows} with
$\delta\geq -30^{\circ}$ (Palla et al.~\cite{palla}, Molinari et 
al.~\cite{mol96},~\cite{mol98a},~\cite{mol00}, Brand et al.~\cite{brand}),
namely that massive protostars are more likely to be found in the 
{\it Low} group. Let us try to shed light on this issue.

\begin{figure}
\centerline{\includegraphics[angle=0,width=9cm]{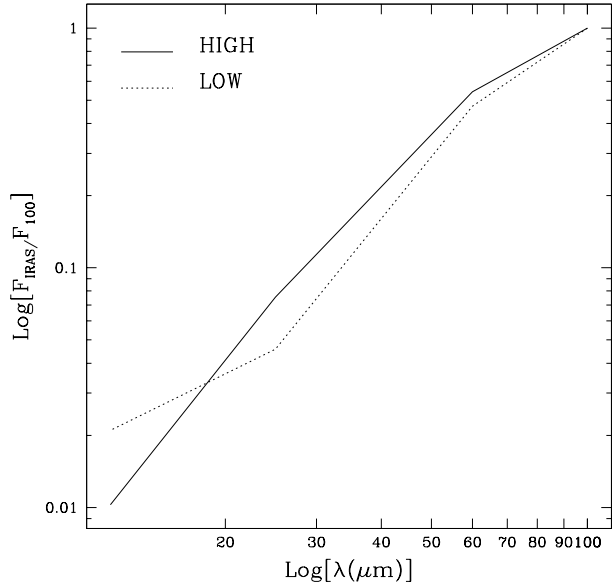}}
\caption[Average IRAS fluxes for {\it High} and {\it Low} sources]{Plot of 
the average IRAS fluxes (normalised for the flux at 100 $\mu$m,
$F_{100}$) for {\it High} (solid line) and {\it Low} (dotted line) sources.}
\label{avspec}
\end{figure}

The distinction between the two groups is basically due to the 
different shape of the SED between 12 and 25 $\mu$m. This is evident 
from Fig.~\ref{avspec}, where we have plotted the average values 
of the IRAS fluxes for 
{\it High} and {\it Low} sources of the sample selected by Palla et 
al.~(\cite{palla}), normalized to the flux at 100 $\mu$m, 
$F_{100}$. Clearly, the average observed 12/100 $\mu$m flux ratio is 
$\sim2$ times larger for {\it Low} sources than for {\it High} sources. 
The emission at this wavelength is due to hot dust; thus a crucial
point concerning
the difference between the two groups is understanding {\it the origin 
of the hot dust}.

Recently, Fontani et al.~(\cite{fonta1}, \cite{fonta2}) have shown
that three {\it Low} sources of the initial sample 
are surrounded by a stellar cluster of more evolved stars.
Given the large beam of the IRAS observations ($\sim30$\asec\ at 12
$\mu$m, which translates into $\sim0.15$ pc at 1 kpc), the IRAS measurements 
at 12 $\mu$m are likely to be significantly affected by the emission of 
such a neighbouring cluster. We indeed concluded that, 
in these three sources, the 
mid-infrared continuum fluxes are dominated by the emission
from the stellar cluster.
Moreover, the presence of a stellar cluster in the surroundings
of the molecular cores has also
been established for a few {\it High} sources (e.g. IRAS 05385+3545, 
Porras et al.~\cite{porras} and IRAS 20126+4104, 
Cesaroni et al.~\cite{cesa97}).
Based upon these results and those of this paper, we
suggest a scenario in which {\it both} {\it Highs} and {\it Lows}
have a high-mass protostellar object embedded in a molecular core, 
and a stellar cluster
located close to the core, but in the {\it Low} sources the flux at 
12$\mu$m is dominated by the emission from the cluster, whereas
the latter is less prominent towards {\it High} sources. 

Note, however, that a nearby
stellar cluster has been studied only in a few sources,
and further observations at high angular resolution
at near- and mid-infrared
wavelengths are absolutely required to support the proposed scenario.

\section{Conclusions}
\label{conc}

We have extended to the southern hemisphere the project started by 
Palla et al.~(\cite{palla}) in the northern sky aimed at identifying 
high-mass protostellar candidates.
From the IRAS-PSC we have selected 131 {\it Low} and 298 
{\it High} sources with $\delta<-30^{\circ}$ using the same criteria
as Palla et al.~(\cite{palla}).
With the aim of testing whether the sources of the {\it Low} group 
are associated with
dense gas, we have observed the CS (2$-$1) and (3$-$2) and \CIII\ (1$-$0)
and (2$-$1) rotational transitions, and the 1.2~mm continuum emission towards
all sources belonging to the {\it Low} group, since
the {\it High} sources had already been observed in CS (2--1)
by Bronfman et al.~(\cite{bronfman}).  
The main findings obtained in this work are:
\begin{itemize}
\item The detection rate of $\sim 85\%$ in CS demonstrates a tight 
association of the sources with dense gas. Among the sources detected
in CS, $\sim 76\%$ have also been detected in 
\CIII\  and $\sim 93\%$ in 1.2~mm continuum.
\item
Continuum maps show the presence of clumps with diameters
in the range $0.2 - 2$ pc and masses from a few $M_{\odot}$ to 
$10^{5}M_{\odot}$; clump kinetic temperatures derived from \CIII\ line ratios
are mostly distributed in the range $\sim 8-10$ K; H$_{2}$ volume densities 
computed from CS line ratios lie between $\sim 10^{4.5}$ and 
$\sim 10^{5.5}$\cmc .
\item
The bolometric luminosities of the sources, derived from IRAS data, are in the
range $10^{3}-10^{6}\;L_{\odot}$, consistent with embedded high-mass objects.
\end{itemize}

Comparing our results to those found in samples of high-mass YSOs with
colours typical of {\it High} sources,
we find that:
\begin{itemize}
\item The luminosities of our sources
are comparable to those found by Sridharan et al.~(\cite{sridharan}) in
their sample of high-mass YSOs. 
This suggests that both samples contain massive YSOs with
comparable masses.
\item The linewidths derived in this 
study are comparable to those observed
in the sample by Sridharan et al.~(\cite{sridharan}), but 
significantly lower than those typically found in clumps associated with
UC \HII\ regions.
Hence, our clumps are less turbulent than those associated with UC \HII\
regions. This can be due to a lower degree of activity of the central objects.
\item The distribution of the NVSS-to-IRAS flux ratios of {\it High} and 
{\it Low} sources of {\it all} luminosities shows that the {\it High} sources
have higher NVSS-to-IRAS flux ratios. The same comparison obtained for
sources with $L<10^{5}L_{\odot}$ shows instead similar distributions
for {\it Highs} and {\it Lows}: 
this indicates that the brightest {\it High}
sources are probably associated with more luminous stars
and/or evolved \HII\ regions.
\item The mass-luminosity ratios found by us are similar to those of
Sridharan et al.~(\cite{sridharan}) but lower than the ratio found 
for a sample 
of UC \HII\ regions. Assuming that clumps of similar masses will form
similar stellar clusters, this result further supports
that our sources, as well as those of Sridharan et al.~(\cite{sridharan}), 
are younger than UC \HII\ regions.
\end{itemize}
Our comparative study suggests that {\it Highs} and {\it Lows} 
with $L<10^{5}L_{\odot}$ are both
associated with massive molecular clumps with similar
physical parameters, indicating that the IRAS colours, on which 
the distinction between the two groups is based, are not indicative of
the relative evolutionary stage.
Based upon these results and those recently achieved by other authors,
we propose that both samples are made of massive clumps hosting high-mass 
protostars, and nearby stellar clusters which are chiefly
responsible for the observed 12~$\mu$m IRAS flux in {\it Low} sources.
Observations with high angular resolution
in the near- and mid-infrared are absolutely required to
confirm this scenario.

\begin{acknowledgements} 
It is a pleasure to thank the ESO/SEST staff for their 
support during the observations.
We thank Robert Zylka (IRAM Grenoble) for helping us with the SIMBA data
reduction, and for his suggestions that improved the quality of the reduction
scripts we used. We also thank the Referee, Dr. Gary Fuller, for
his useful suggestions and comments.
\end{acknowledgements}

{}
 
\section{tables}
\begin{table*}[h]
\begin{center}
\caption[] {Observed sources and detection summary. R.A.(J2000) and
Dec.(J2000) represent the equatorial coordinates of the IRAS source.
$v_{\rm LSR}$ is the
velocity at which we centered the CS spectra during the two
observing runs (see text). 
N.O. means that the source was not observed in that tracer.
$\Delta$ is the angular separation between the IRAS position and the
nearest millimeter peak detected in the SIMBA maps.}
\label{tsources}
\begin{tabular}{cccccccccc}
\hline
\small
IRAS name  & R.A.(J2000) & Dec.(J2000) & \multicolumn{2}{c}{$v_{\rm LSR}$} & CS & \CIII\
 & 1.2~mm  & $\Delta$ \\
      &             &            & \multicolumn{2}{c}{(\kms )} & & & & (\asec ) \\
      &             &            & run I & run II & & & & \\ 
\hline
08211$-$4158 & 08:22:52.3 & $-$42:07:57 &   16.0 & $-$ &  Y  &    Y  & Y & 10 \\
08247$-$4223 & 08:26:27.6 & $-$42:33:05 &   16.1 & $-$  &  Y  &    N  & Y & 40 \\
08477$-$4359 & 08:49:32.9 & $-$44:10:47 &   70.6 & $-$ &  Y  &    Y  & Y & 30 \\
08488$-$4457 & 08:50:38.2 & $-$45:08:18 &   62.1 & $-$ &  N  &    N.O. & $\spadesuit$ & $-$ \\
08563$-$4225 & 08:58:12.5 & $-$42:37:34 &   20.9 & $-$  &  Y  &    Y & Y & 15  \\
09014$-$4736 & 09:03:09.8 & $-$47:48:28 &    0.0 & $-$ &  Y  &    N  & Y & 5 \\
09026$-$4842 & 09:04:22.2 & $-$48:54:21 &    0.0 & $-$ &  Y  &    Y  & Y & 16 \\
09131$-$4723 & 09:14:55.5 & $-$47:36:13 &    0.0 & $-$ &  Y  &    Y  & Y & 30 \\
09166$-$4813 & 09:18:26.6 & $-$48:26:26 &    0.0 & $-$ &  Y  &    Y  & Y & 80 \\
09209$-$5143 & 09:22:34.6 & $-$51:56:23 &    0.0 & $-$ &  Y  &    Y  & Y & 5 \\
10088$-$5730 & 10:10:38.7 & $-$57:45:32 &   18.8 & $-$ &  N  &    N.O. & $\spadesuit$ & $-$ \\ 
10095$-$5843 & 10:11:15.8 & $-$58:58:15 &    0.0 & $-$ &  Y  &    Y  & Y & 15 \\
10102$-$5706 & 10:12:03.7 & $-$57:21:26 &    0.0 & $-$ &  Y  &    N  & N & $-$ \\
10123$-$5727 & 10:14:08.8 & $-$57:42:12 &    0.0 & $-$ &  Y  &    Y  & Y & 30 \\
10156$-$5804 & 10:17:26.8 & $-$58:19:46 &    0.0 & $-$ &  N  &    N  & N & $-$ \\
10277$-$5730 & 10:29:35.4 & $-$57:45:34 &    0.0 & $-$ &  Y  &    N  & Y & 25 \\
10308$-$6122 & 10:32:39.8 & $-$61:37:33 &   $-$7.3 & $-$ &  Y  &    N  & Y & 10 \\
10317$-$5936 & 10:33:38.1 & $-$59:51:54 &    0.0 & $-$ &  Y  &    N  & Y & 25 \\
10439$-$5941 & 10:45:54.0 & $-$59:57:03 &    0.0 & $-$ &  Y  &    Y  & Y & 15 \\
10521$-$6031 & 10:54:11.0 & $-$60:47:30 &    0.0 & $-$ &  Y  &    N  & Y & 25 \\
10537$-$5930 & 10:55:49.0 & $-$59:46:47 &    0.0 & $-$ &  Y  &    N  & Y & 20 \\
10545$-$6244 & 10:56:32.9 & $-$63:00:34 &    0.0 & $-$ &  Y  &    N  & N & $-$ \\
10548$-$5929 & 10:56:51.9 & $-$59:45:14 &    0.0 & $-$ &  Y  &    N  & Y & 15 \\
10554$-$6237 & 10:57:25.0 & $-$62:53:10 &    0.0 & $-$ &  Y  &    Y  & Y & 10 \\
10555$-$5949 & 10:57:37.5 & $-$60:05:32 &    0.0 & $-$ &  Y  &   N.O. & N.O. & $-$ \\
10572$-$6018 & 10:59:19.3 & $-$60:34:10 &    0.0 & $-$ &  Y  &    N  & Y & 40 \\
10575$-$5844 & 10:59:40.3 & $-$59:01:05 &    0.0 & $-$ &  N  &    N  & N & $-$  \\
10591$-$5934 & 11:01:15.8 & $-$59:51:01 &    0.0 & $-$ &  Y  &    N  & Y & 30 \\
11265$-$6158 & 11:28:50.9 & $-$62:15:01 &    0.0 & $-$ &  Y  &    N  & Y & 15 \\
11294$-$6257 & 11:31:46.5 & $-$63:14:25 &    0.0 & $-$ &  Y  &    N  & Y & 15 \\
11380$-$6311 & 11:40:27.6 & $-$63:27:56 &    0.0 & $-$ &  Y  &    Y  & Y & 25 \\
11396$-$6202 & 11:42:01.5 & $-$62:19:24 &    0.0 & $-$ &  Y  &    N  & Y & 60 \\
11404$-$6215 & 11:42:48.0 & $-$62:32:20 &   39.0 & $-$ &  Y  &    Y  & Y & 40 \\
11476$-$6435 & 11:50:08.0 & $-$64:52:20 &    0.0 & $-$ &  N  &    N.O. & $\spadesuit$ & $-$ \\
\hline
\end{tabular}
\end{center}
\end{table*}
\addtocounter{table}{-1}
\begin{table*}
\caption[] {Continued. }
\begin{center}
\begin{tabular}{ccccccccc}
\hline
IRAS name  & R.A.(J2000) & Dec.(J2000) & \multicolumn{2}{c}{$v_{\rm LSR}$} & CS & \CIII\
 & 1.2~mm  & $\Delta$ \\
      &             &            & \multicolumn{2}{c}{(\kms )} & & & & (\asec ) \\
      &             &            & run I & run II & & & & \\
\hline
12102$-$6133 & 12:12:57.9 & $-$61:50:17 &    0.0 & $-$ &  Y  &    N & Y & 50 \\ 
12295$-$6224 & 12:32:22.7 & $-$62:41:25 &    0.0 & $-$ &  Y  &    Y & Y & 15 \\
12377$-$6237 & 12:40:42.4 & $-$62:54:09 &    0.0 & $-$ &  Y  &    N & Y & 0 \\
12434$-$6355 & 12:46:24.6 & $-$64:11:26 &    0.0 & $-$  &  N  &    N.O. & $\spadesuit$ & $-$ \\
13023$-$6213 & 13:05:30.7 & $-$62:29:58 &  $-$10.0 & $-$ &  Y  &    Y & Y & 15 \\
13039$-$6108 & 13:07:07.0 & $-$61:24:47 &  $-$30.0 & $-$ &  Y  &    Y & Y & 22 \\
13078$-$6247 & 13:11:05.1 & $-$63:03:48 &    0.0 & $-$ &  N  &    N.O. & $\spadesuit$ & $-$ \\
13106$-$6050 & 13:13:50.5 & $-$61:06:44 &    0.0 & $-$ &  Y  &    Y & Y & 8 \\
13333$-$6234 & 13:36:45.7 & $-$62:49:36 &  $-$30.0 & $-$ &  Y  &    Y & Y & 16 \\
13384$-$6152 & 13:41:53.3 & $-$62:07:36 &    0.0 & $-$ &  Y  &    Y & Y & 10 \\
13395$-$6153 & 13:42:59.5 & $-$62:08:43 &    0.0 & $-$ &  Y  &    Y & Y & 16 \\
13438$-$6203 & 13:47:21.9 & $-$62:18:41 &  $-$30.0 & $-$ &  Y  &    Y & Y & 40 \\
13481$-$6124 & 13:51:37.8 & $-$61:39:08 &    0.0 & $-$ &  Y  &    Y & Y & 0 \\
13558$-$6159 & 13:59:27.0 & $-$62:13:40 &  $-$30.0 & $-$45.0  & N & N.O. & $\spadesuit$ & $-$ \\
13560$-$6133 & 13:59:35.3 & $-$61:48:17 &    0.0 & $-$ & Y & Y  & Y & 40 \\
14000$-$6104 & 14:03:36.6 & $-$61:18:28 &    0.0 & $-$ & Y & Y  & Y & 0 \\
14131$-$6126 & 14:16:48.6 & $-$61:40:26 &  $-$30.0 & $-$ & Y & Y  & Y & 20 \\
14166$-$6118 & 14:20:19.5 & $-$61:31:53 &  $-$30.0 & $-$ & Y & Y  & Y & 0 \\
14183$-$6050 & 14:22:02.8 & $-$61:04:18 &    0.0 & $-$ & Y & N  & Y & 10 \\
14198$-$6115 & 14:23:33.2 & $-$61:28:53 &    0.0 & $-$ & N & N.O. & $\spadesuit$ & $-$ \\
14201$-$6044 & 14:23:54.2 & $-$60:57:45 &  $-$30.0 & $-$ & Y & Y & Y & 20 \\
14395$-$5941 & 14:43:20.9 & $-$59:53:54 &    0.0 & $-$ & Y & Y & Y & 25 \\
14412$-$5948 & 14:45:04.1 & $-$60:01:14 &    0.0 & $-$ & N & N.O. & $\spadesuit$ & $-$  \\
14425$-$6023 & 14:46:23.5 & $-$60:35:45 &    0.0 & $-$ & Y & Y & Y & 15 \\
14591$-$5843 & 15:02:58.8 & $-$58:55:06 &  $-$40.0 & $-$ & Y & N & Y & 40 \\
15038$-$5828 & 15:07:43.8 & $-$58:39:53 &    0.0 & $-$ & Y & Y & Y & 100 \\
15072$-$5855 & 15:11:07.9 & $-$59:06:30 &  $-$20.0 & $-$ & Y & Y & Y & 18 \\
15100$-$5903 & 15:14:00.0 & $-$59:15:09 &  $-$40.0 & $-$ & Y & Y & Y & 36 \\
15178$-$5641 & 15:21:45.4 & $-$56:52:42 &  $-$30.0 & $-$ & Y & N & Y & 15  \\
15219$-$5658 & 15:25:48.7 & $-$57:09:11 &  $-$20.0 & $-$ & Y & Y & Y & 15 \\
15239$-$5538 & 15:27:49.3 & $-$55:48:42 &  $-$40.0 & $-$ & Y & N & Y & 40 \\
15246$-$5612 & 15:28:32.6 & $-$56:23:00 &  $-$20.0 & $-$ & Y & Y & Y & 15 \\
15262$-$5541 & 15:30:05.5 & $-$55:52:01 &  $-$20.0 & $-$55.7 & Y & Y & Y & 40 \\
\hline
\end{tabular}
\end{center}
\end{table*}
\addtocounter{table}{-1}
\begin{table*}
\caption[] {Continued. }
\begin{center}
\begin{tabular}{ccccccccc}
\hline
IRAS name  & R.A.(J2000) & Dec.(J2000) & \multicolumn{2}{c}{$v_{\rm LSR}$} & CS & \CIII\
 & 1.2~mm  & $\Delta$ \\
      &             &            & \multicolumn{2}{c}{(\kms )} & & & & (\asec ) \\
      &             &            & run I & run II & & & & \\
\hline
15347$-$5518 & 15:38:36.0 & $-$55:28:07 &  $-$40.0 & $-$ & Y & Y & Y & 0 \\
15371$-$5458 & 15:40:58.6 & $-$55:08:20 &  30.0 & $-$ & Y & N & Y & 0  \\
15470$-$5419 & 15:50:55.2 & $-$54:28:22 &  $-$20.0 & $-$ & Y & Y & Y & 33 \\
15506$-$5325 & 15:54:32.2 & $-$53:33:53 &  $-$50.0 & $-$ & N & N.O. & $\spadesuit$ & $-$ \\
15519$-$5430 & 15:55:50.4 & $-$54:38:58 &  $-$10.0 & $-$ & Y & Y & Y & 34 \\
15557$-$5337 & 15:59:38.2 & $-$53:45:32 &  $-$40.0 & $-$ & Y & Y & Y & 30 \\
15571$-$5218 & 16:01:00.4 & $-$52:27:13 &  $-$50.0 & $-$ & Y & N & N & $-$ \\
15579$-$5303 & 16:01:46.6 & $-$53:11:41 &  $-$20.0 & $-$ & Y & Y & Y & 0 \\
15579$-$5347 & 16:01:52.6 & $-$53:56:21 &  $-$20.0 & $-$ & N.O. & N.O. & $\spadesuit$ & $-$  \\
15583$-$5314 & 16:02:10.1 & $-$53:22:35 &  $-$50.0 & $-$ & Y & Y & Y & 15 \\
16061$-$5048 & 16:09:57.3 & $-$50:56:45 &  $-$50.0 & $-$ & Y & Y & Y & 25 \\
16082$-$5031 & 16:12:03.2 & $-$50:39:16 &  $-$32.4 & $-$ & Y & Y & Y & 10 \\
16093$-$5015 & 16:13:05.2 & $-$50:23:05 &  $-$60.0 & $-$ & Y & Y & Y & 40 \\
16093$-$5128 & 16:13:09.2 & $-$51:36:26 &  $-$50.0 & $-$ & Y & Y & Y & 25 \\
16106$-$5048 & 16:14:26.8 & $-$50:56:12 &  $-$21.5 & $-$ & Y & Y & Y & ?  \\
16107$-$4956 & 16:14:29.4 & $-$50:03:51 &  $-$60.0 & $-$ & Y & Y & Y & 20 \\
16148$-$5011 & 16:18:35.2 & $-$50:18:53 &  $-$60.0 & $-$ & Y & Y & Y & 0 \\
16153$-$5016 & 16:19:07.0 & $-$50:24:12 &  $-$23.4 & $-$ & Y & Y & Y & 100 \\
16170$-$5053 & 16:20:53.2 & $-$51:00:14 &  $-$60.0 & $-$54.0 & Y & N & Y & 50 \\
16187$-$4932 & 16:22:29.3 & $-$49:39:02 &  $-$60.0 & $-$ & Y & Y & Y & 38 \\  
16194$-$4934 & 16:23:13.2 & $-$49:40:59 &  $-$25.2 & $-$ & Y & Y & Y & 35 \\
16204$-$4916 & 16:24:12.3 & $-$49:23:34 &  $-$60.0 & $-$ & Y & Y & Y & 20 \\
16204$-$4943 & 16:24:15.5 & $-$49:50:05 &  $-$60.0 & 0.0 & N & N.O. & $\spadesuit$ & $-$ \\
16218$-$4931 & 16:25:37.9 & $-$49:38:20 &  $-$50.0 & $-$  & Y & Y & Y & 0 \\
16219$-$4848 & 16:25:39.4 & $-$48:55:12 &  $-$60.0 & $-$ & Y & Y & N & $-$ \\
16231$-$4819 & 16:26:49.1 & $-$48:25:49 &  $-$60.0 & $-$ & Y & N & N.O. & $-$ \\
16232$-$4917 & 16:27:02.0 & $-$49:23:52 &  $-$50.0 & $-$ & Y & Y & Y & 20 \\
16252$-$4853 & 16:29:01.6 & $-$48:59:48 &  $-$60.0 & $-$ & Y & Y & Y & 16 \\
16254$-$4844 & 16:29:09.0 & $-$48:51:27 &  $-$60.0 & $-$ & Y & Y & Y & 80 \\
16344$-$4605 & 16:38:08.6 & $-$46:11:10 &  $-$50.0 & $-$ & Y & Y & Y & 15 \\
16358$-$4614 & 16:39:29.4 & $-$46:20:44 &  $-$60.0 & $-$ & N & N & N.O. & $-$  \\
\hline
\end{tabular}
\end{center}
\end{table*}
\addtocounter{table}{-1}
\begin{table*}
\caption[] {Continued. }
\begin{center}
\begin{tabular}{ccccccccc}
\hline
\small
IRAS name  & R.A.(J2000) & Dec.(J2000) & \multicolumn{2}{c}{$v_{\rm LSR}$} & CS & \CIII\
 & 1.2~mm  & $\Delta$ \\
      &             &            & \multicolumn{2}{c}{(\kms )} & & & & (\asec ) \\
      &             &            & run I & run II & & & & \\
\hline
16363$-$4645 & 16:40:00.5 & $-$46:51:32 &  $-$60.0 & $-$ & Y & Y & Y & 25 \\
16369$-$4810 & 16:40:38.2 & $-$48:16:00 &  $-$60.0 & $-$ & Y & Y & Y & ? \\
16403$-$4614 & 16:44:01.5 & $-$46:20:27 &  $-$50.0 & $-$ & Y & Y & N & $-$  \\
16404$-$4518 & 16:44:05.9 & $-$45:23:53 &  $-$60.0 & $-$ & N & N & N.O. & $-$ \\
16417$-$4445 & 16:45:19.9 & $-$44:51:00 &  $-$60.0 & $-$ & Y & Y & N & $-$  \\
16419$-$4602 & 16:45:37.9 & $-$46:07:49 &  $-$ & $-$60.0 & Y & Y & Y & 25 \\
16428$-$4109 & 16:46:22.6 & $-$41:14:58 &  $-$60.0 & $-$ & Y & Y & N & $-$ \\
16464$-$4359 & 16:50:01.0 & $-$44:05:03 &  $-$50.0 & $-$ & Y & Y & Y & 10  \\
16501$-$4314 & 16:53:41.1 & $-$43:19:23 &  $-$50.0 & $-$ & Y & Y & Y & ? \\
16535$-$4300 & 16:57:05.5 & $-$43:05:20 &  $-$60.0 & $-$ & Y & Y & Y & 20 \\
16573$-$4214 & 17:00:54.3 & $-$42:19:10 &  $-$ & $-$25.0 & Y & Y & Y & ? \\
16581$-$4212 & 17:01:38.8 & $-$42:17:05 &  $-$90.0 & 0.0 & N & N.O. & $\spadesuit$ & $-$ \\
17033$-$4035 & 17:06:49.3 & $-$40:39:51 &  $-$60.0 & $-$ & Y & Y & Y & 30 \\
17036$-$4033 & 17:07:08.9 & $-$40:37:08 &  $-$60.0 & $-$ & Y & Y & Y & 10 \\
17040$-$3959 & 17:07:33.7 & $-$40:03:04 &  0.0 & $-$ & Y & Y & Y & 15 \\
17082$-$4114 & 17:11:46.2 & $-$41:18:03 &  $-$ & $-$40.0 & Y & Y & Y & 40 \\
17114$-$3804 & 17:14:52.2 & $-$38:07:26 &  $-$30.0 & $-$ & N & N & N.O. & $-$ \\
17140$-$3747 & 17:17:28.1 & $-$37:51:06 &  $-$30.0 & $-$ & N & N & N & $-$  \\
17141$-$3606 & 17:17:28.2 & $-$36:09:38 &  0.0 & $-$ & Y & N & Y & 30 \\
17156$-$3607 & 17:19:01.2 & $-$36:10:12 &  $-$30.0 & $-$ & Y & Y & N & $-$  \\
17211$-$3537 & 17:24:28.5 & $-$35:40:13 &  $-$20.0 & $-$ & Y & Y & Y & 4 \\
17218$-$3704 & 17:25:14.0 & $-$37:07:29 &  $-$50.0 & $-$ & Y & Y & Y & 25 \\
17225$-$3426 & 17:25:50.3 & $-$34:29:21 &  $-$30.0 & $-$ & Y & Y & Y & ? \\
17230$-$3531 & 17:26:26.3 & $-$35:33:35 &  $-$20.0 & $-$ & Y & Y & Y & 5 \\
17256$-$3631 & 17:29:01.1 & $-$36:33:38 &  $-$20.0 & $-$ & Y & Y & Y & 25 \\
17285$-$3346 & 17:31:48.1 & $-$33:48:23 &   0.0 & $-$ & Y & Y & Y & 12 \\
17338$-$3044 & 17:37:03.9 & $-$30:46:17 &  0.0 & $-$ & Y & Y & Y & 30 \\
17355$-$3241 & 17:38:50.5 & $-$32:43:35 &  0.0 & $-$ & Y & Y & Y & 5 \\
17368$-$3057 & 17:40:05.8 & $-$30:58:51 &  0.0 & $-$ & N & N & N.O. & $-$ \\
17377$-$3109 & 17:40:57.2 & $-$31:11:00 &  0.0 & $-$ & Y & Y & Y & 10 \\
17410$-$3019 & 17:44:14.9 & $-$30:20:42 &  0.0 & $-$ & Y & Y & Y & 25 \\
17419$-$3207 & 17:45:10.4 & $-$32:08:48 &  0.0 & $-$ & N & N & N.O. & $-$  \\
17425$-$3017 & 17:45:45.1 & $-$30:18:51 &  0.0 & $-$ & Y & Y & Y & 20 \\
\hline
\normalsize
\end{tabular}
\end{center}
$^{\spadesuit}$ the continuum maps will be available in 
a forthcoming paper (Beltran et al., in prep.) \\
\end{table*}

\begin{table*}
\caption[] {CS line parameters obtained from Gaussian fits$^{(\diamondsuit)}$ 
. Typical rms noise in the spectra is $\sim 0.05$ K and $0.06$ K
for the (2--1) and (3--2) lines, respectively.}
\label{tcsline}
\begin{center}
\begin{tabular}{ccccccc}
\hline
source & & CS(2$-$1) & & & CS(3$-$2)& \\
\hline
       &$\int T_{\rm MB}{\rm d}v$ & $v_{\rm LSR}$ & FWHM & $\int T_{\rm MB}{\rm d}v$ & $v_{\rm LSR}$ & FWHM \\
       & (K \kms ) & (\kms ) & (\kms ) & (K \kms ) & (\kms ) & (\kms )  \\
\hline
08211$-$4158 & 2.61(0.03) & 10.80 & 1.76 & 1.99 & 10.93 & 1.59 \\
08247$-$4223 & 0.42 & 8.04 & 1.5 & 0.38 & 7.81 & 1.9 \\
08477$-$4359 & 2.27 & 8.47 & 2.97 & 1.82 & 8.15 & 2.76 \\
08488$-$4457 & $\leq 0.36$ & $-$ & $-$ & $\leq 0.30$ & $-$ & $-$ \\
08563$-$4225 & 9.22 & 7.66 & 3.78 & 9.51 & 7.67 & 3.54 \\
09014$-$4736 & 0.39 & 3.26 & 2.14 & $\leq 0.30$ & $-$ & $-$ \\
09026$-$4842 & 2.65 & 5.15 & 1.83 & 2.06 & 5.26 & 1.68 \\
09131$-$4723 & 3.74 & 4.03 & 2.43 & 2.72 & 4.03 & 2.62 \\
09166$-$4813 & 0.82 & 6.28 & 1.9 & 0.29 & 6.47 & 1.19 \\
09209$-$5143 & 0.97 & 36.85 & 1.89 & $\leq 0.33$ & $-$ & $-$ \\
10088$-$5730 & $\leq 0.34$ & $-$ & $-$ & $\leq 0.33$ & $-$ & $-$ \\
10095$-$5843 & 3.62 & $-$4.57 & 2.40 & 3.16 & $-$4.38 & 1.98 \\
10102$-$5706$^{\spadesuit}$ & 0.48 & $-$3.59($-$7,0) & 2.12 & $\leq 0.33$ & $-$ & $-$ \\
10123$-$5727$^{\spadesuit}$ & 4.60 & $-$4.03($-$7,$-$1) & 2.73 & 2.39 & $-$3.87($-$7,$-$1) & 2.47 \\
10156$-$5804 & $\leq 0.25$ & $-$ & $-$ & $\leq 0.25$ & $-$ & $-$ \\
10277$-$5730 & 1.10 & 10.65 & 2.5 & 1.23 & 10.95 & 2.68 \\
10308$-$6122 & 0.72 & $-$6.65 & 2.32 & 0.52 & $-$6.90 & 2.28 \\
10317$-$5936 & 0.62 & 37.35 & 3.47 & $\leq 0.33$ & $-$ & $-$ \\
10439$-$5941 & 5.34 & $-$14.57 & 2.42 & 4.33 & $-$14.54 & 2.01 \\
10521$-$6031 & 0.87 & 23.81 & 2.3 & 0.48 & 23.88 & 2.0 \\
10537$-$5930 & 0.68 & 15.38 & 3.7 & $\leq 0.33$ & $-$ & $-$ \\
10545$-$6244 & 0.39 & $-$11.62 & 1.9 & 0.31 & $-$11.21 & 1.47 \\
10548$-$5929 & 0.75 & 18.85 & 2.7 & 0.56 & 19.16 & 2.23 \\
10554$-$6237 & 3.58 & $-$16.78 & 2.18 & 2.61 & $-$16.69 & 2.07 \\
10555$-$5949 & $\leq 0.31$ & $-$ & $-$ & 0.49 & 28.08 & 2.1 \\
10572$-$6018 & 0.93 & 13.09 & 2.6 & 0.75 & 13.19 & 1.89 \\
10575$-$5844 & $\leq 0.33$ & $-$ & $-$ & $\leq 0.33$ & $-$ & $-$ \\
10591$-$5934 & 1.15 & $-$25.92 & 2.1 & 0.59 & $-$25.47 & 2.4 \\
11265$-$6158 & 1.42 & $-$23.27 & 1.96 & 1.12 & $-$23.21 & 1.76 \\
11294$-$6257 & 1.04 & $-$26.13 & 2.0 & $\leq 0.36$ & $-$ & $-$ \\
11380$-$6311 & 4.06 & $-$11.33 & 2.71 & 2.67 & $-$11.09 & 2.70 \\
11396$-$6202 & 0.62 & 41.07 & 2.38 & $\leq 0.33$ & $-$ & $-$ \\
11404$-$6215 & 1.77 & 38.86 & 2.9 & 1.78 & 39.07 & 2.75 \\
11476$-$6435 & $\leq 0.30$ & $-$ & $-$ & $\leq 0.36$ & $-$ & $-$ \\
12102$-$6133 & 0.54 & $-$32.58 & 1.6 & 0.27 & $-$32.41 & 0.9 \\
12295$-$6224 & 3.51 & $-$36.83 & 2.21 & 2.44 & $-$36.48 & 2.11 \\
12377$-$6237 & 0.44 & 22.0 & 2.8 & $\leq 0.36$ & $-$ & $-$ \\
\hline
\end{tabular}
\end{center}
\end{table*}
\addtocounter{table}{-1}
\begin{table*}
\caption[] {Continued. }
\begin{center}
\begin{tabular}{ccccccc}
\hline
source & & CS(2$-$1) & & & CS(3$-$2)& \\
\hline
       &$\int T_{\rm MB}{\rm d}v$ & $v_{\rm LSR}$ & FWHM & $\int T_{\rm MB}{\rm d}v$ & $v_{\rm LSR}$ & FWHM \\
       & (K \kms ) & (\kms ) & (\kms ) & (K \kms ) & (\kms ) & (\kms )  \\
\hline
12434$-$6355 & $\leq 0.36$ & $-$ & $-$ & $\leq 0.33$ & $-$ & $-$ \\
13023$-$6213 & 3.22 & $-$43.19 & 2.82 & 2.55 & $-$42.99 & 2.46 \\
13039$-$6108 & 1.65 & $-$26.21 & 1.70 & 0.95 & $-$26.15 & 1.39 \\
13078$-$6247 & $\leq 0.33$ & $-$ & $-$ & $\leq 0.33$ & $-$ & $-$ \\
13106$-$6050 & 3.42 & $-$57.32 & 2.37 & 1.90 & $-$57.42 & 2.24 \\
13333$-$6234 & 8.66 & $-$11.44 & 4.54 & 7.13 & $-$11.14 & 3.73 \\
13384$-$6152 & 2.75 & $-$50.44 & 2.57 & 1.53 & $-$50.36 & 2.04 \\
13395$-$6153 & 18.04 & $-$50. 84 & 3.83 & 17.52 & $-$50.75 & 3.69 \\
13438$-$6203$^{\spadesuit}$ & 3.27 & $-$51.27($-$58,$-$46) & 6.03 & 0.43 & $-$51.10 & 2.1 \\
13481$-$6124 & 8.12 & $-$41.38 & 3.13 & 5.05 & $-$40.98 & 2.68 \\
13558$-$6159 & $\leq 0.35$ & $-$ & $-$ & $\leq 0.36$ & $-$ & $-$ \\
13560$-$6133 & 3.38 & $-$58.37 & 4.05 & 3.18 & $-$58.56 & 3.44 \\
14000$-$6104 & 3.53 & $-$58.92 & 4.3 & 4.74 & $-$58.95 & 4.24 \\
14131$-$6126 & 0.77 & $-$34.54 & 1.34 & 0.84 & $-$34.42 & 1.44 \\
14166$-$6118 & 1.63 & $-$41.12 & 2.96 & 0.94 & $-$41.05 & 2.67 \\
14183$-$6050 & 0.60 & $-$42.64 & 1.10 & 1.13 & $-$42.65 & 1.43 \\
14198$-$6115 & $\leq 0.36$ & $-$ & $-$ & $\leq 0.36$ & $-$ & $-$ \\
14201$-$6044 & 0.94 & $-$49.84 & 1.89 & 0.94 & $-$49.65 & 1.93 \\
14395$-$5941 & 1.53 & $-$42.40 & 2.22 & 0.98 & $-$42.25 & 3.17 \\
14412$-$5948 & $\leq 0.38$ & $-$ & $-$ & $\leq 0.36$ & $-$ & $-$ \\
14425$-$6023 & 5.97 & $-$45.89 & 3.13 & 5.24 & $-$45.74 & 3.19 \\
14591$-$5843 & 1.14 & $-$29.69 & 4.1 & $\leq 0.33$ & $-$ & $-$ \\
15038$-$5828 & 1.64 & $-$67.14 & 3.65 & $\leq 0.35$ & $-$ & $-$ \\
15072$-$5855 & 4.73 & $-$41.74 & 2.19 & 4.11 & $-$41.65 & 2.23 \\
15100$-$5903 & 1.79 & $-$51.27 & 2.39 & 0.70 & $-$50.91 & 2.77 \\
15178$-$5641 & 0.90 & $-$28.47 & 4.8 & $\leq 0.33$ & $-$ & $-$ \\
15219$-$5658 & 2.93 & $-$15.99 & 4.0 & 2.99 & $-$16.05 & 4.12 \\ 
15239$-$5538 & 0.76 & $-$46.79 & 2.5 & 0.41  & $-$46.44  & 3.6  \\
15246$-$5612 & 5.50 & $-$65.44 & 3.06 & 3.57 & $-$65.35 & 2.65 \\ 
15262$-$5541 & 1.00 & $-$54.15 & 2.9 & 0.17 & $-$54.61 & 0.96 \\
15347$-$5518 & 3.77 & $-$61.54 & 3.04 & 2.59 & $-$61.36 & 2.88 \\
15371$-$5458 & 0.62 & 31.98 & 3.84 & 0.65 & 32.3  & 4.8 \\
15470$-$5419 & 3.83 & $-$61.73 & 4.9 & 3.11 & $-$61.43 & 3.97 \\
15506$-$5325 & $\leq 0.32$ & $-$ & $-$ & $\leq 0.35$ & $-$ & $-$ \\
15519$-$5430 & 10.92 & $-$36.53 & 2.87 & 8.34 & $-$36.40 & 2.72 \\
15557$-$5337 & 46.28 & $-$47.12 & 4.21 & 56.82 & $-$47.02 & 4.51 \\
15571$-$5218$^{\spadesuit}$ & 1.40 & $-$101.3($-$107,$-$98) & 5.36 & 0.70 & $-$100.7($-$107,$-$98) & 4.84 \\
\hline
\end{tabular}
\end{center}
\end{table*}
\addtocounter{table}{-1}
\begin{table*}
\caption[] {Continued. }
\begin{center}
\begin{tabular}{ccccccc}
\hline
source & & CS(2$-$1) & & & CS(3$-$2)& \\
\hline
       &$\int T_{\rm MB}{\rm d}v$ & $v_{\rm LSR}$ & FWHM & $\int T_{\rm MB}{\rm d}v$ & $v_{\rm LSR}$ & FWHM \\
       & (K \kms ) & (\kms ) & (\kms ) & (K \kms ) & (\kms ) & (\kms )  \\
\hline
15579$-$5303$^{\spadesuit}$ & 15.2 & $-$49.8($-$60,$-$34) & 11.87 & 14.45 & $-$49.25($-$62,$-$34) & 11.60 \\
15583$-$5314 & 2.57 & $-$77.58 & 2.44 & 1.08 & $-$77.30 & 2.15 \\
16061$-$5048 & 3.72 & $-$51.81 & 2.88 & 1.71 & $-$51.67 & 2.07 \\
16082$-$5031 & 2.10 & $-$41.00 & 2.31 & 1.51 & $-$40.95 & 2.26 \\
16093$-$5015 & 2.43 & $-$42.93 & 2.04 & 1.28 & $-$42.82 & 1.66 \\
16093$-$5128 & 2.29 & $-$97.32 & 3.6 & 1.47 & $-$96.96 & 3.45 \\
16106$-$5048$^{\spadesuit}$& 1.99 & $-$87.55($-$92,$-$80) & 5.96 & 0.95 & $-$87.36($-$92,$-$80) & 5.87 \\
16107$-$4956 & 0.69 & $-$83.20 & 2.0 & 0.32 & $-$82.89 & 1.45 \\
16148$-$5011 & 5.22 & $-$44.76 & 2.64 & 4.04 & $-$44.47 & 2.02 \\
16153$-$5016 & 0.58 & $-$41.3 & 2.5 & 0.40 & $-$41.44 & 1.99 \\
16170$-$5053$^{\spadesuit}$ & 2.53 & $-$54.2($-$59,$-$49) & 4.25 & 0.55 & $-$53.98 & 2.78 \\
16187$-$4932 & 1.50 & $-$60.58 & 3.56 & 0.32 & $-$59.89 & 1.77 \\
           &$^{\clubsuit}$ 0.74 & $-$48.27 & 2.69 & $\leq 0.30$ & $-$ & $-$ \\
           & 0.48 & $-$44.65 & 1.74 & $\leq 0.30$ & $-$ & $-$ \\
16194$-$4934 & 1.41 & $-$85.27 & 3.53 & 0.56 & $-$84.64 & 2.67 \\
16204$-$4916$^{\spadesuit}$ & 8.34 & $-$70.43($-$85,$-$60) & 10.02 & 4.29 & $-$70.60($-$85,$-$62) & 10.35 \\
16204$-$4943 & $\leq 0.54$ & $-$ & $-$ & $\leq 0.38$ & $-$ & $-$ \\
16218$-$4931$^{\spadesuit}$ & 3.33 & $-$37.80($-$45,$-$30) & 6.32 & 2.85 & $-$37.64($-$45,$-$30) & 5.1 \\
16219$-$4848 & 0.95 & $-$79.99 & 2.5 & 0.74 & $-$79.81 & 2.89 \\
16231$-$4819 & 0.42 & $-$59.1  & 2.4  & $\leq 0.34$ & $-$ & $-$ \\
16232$-$4917 & 3.22 & $-$46.41 & 2.13 & 2.36 & $-$46.37 & 1.99 \\
16252$-$4853 & 1.19 & $-$45.87 & 1.72 & 0.43 & $-$45.77 & 1.65 \\
16254$-$4844 & 1.55 & $-$45.56 & 3.03 & 0.56 & $-$45.27 & 2.44 \\
           &$^{\clubsuit}$ 0.93 & $-$40.84 & 1.49 & 0.80 & $-$40.85 & 1.51 \\
16344$-$4605 &$^{\clubsuit}$ 5.93 & $-$61.53 & 5.25 & 0.80 & $-$63.95 & 2.25 \\
           & $\leq 0.40$ &  & & 2.95 & $-$60.64 & 3.54 \\
16358$-$4614 & $\leq 0.36$ & $-$ & $-$ & $\leq 0.36$ & $-$ & $-$ \\
16363$-$4645 & 2.09 & $-$65.29 & 3.39 & 1.64 & $-$65.23 & 2.49 \\
           & 0.29 & $-$61.02 & 2.12 & 0.90 & $-$61.85 & 3.11 \\
16369$-$4810 & 3.48 & $-$39.09 & 4.58 & 1.44 & $-$39.31 & 3.65 \\
16403$-$4614$^{\spadesuit}$ & 0.95 & $-$120.1($-$124,$-$117) & 3.62 & $\leq 0.37$ & $-$ & $-$ \\
16404$-$4518 & $\leq 0.42$ & $-$ & $-$ & $\leq 0.37$ & $-$ & $-$ \\
16417$-$4445 & 1.14 & $-$56.06 & 2.2 & 0.78 & $-$56.13 & 2.0 \\
16419$-$4602 & 2.22 & $-$37.29 & 2.68 & $\leq 0.75$$^{*}$ & $-$ & $-$ \\
16428$-$4109 & 0.77 & $-$25.65 & 1.21 & 0.50 & $-$25.80 & 1.02 \\
16464$-$4359 & 2.55 & $-$79.09 & 3.14 & 1.83 & $-$79.44 & 2.59 \\
16501$-$4314 & 5.63 & $-$118.96 & 3.10 & 4.13 & $-$118.79 & 2.68 \\
\hline
\end{tabular}
\end{center}
\end{table*}
\addtocounter{table}{-1}
\begin{table*}
\caption[] {Continued. }
\begin{center}
\begin{tabular}{ccccccc}
\hline
source & & CS(2$-$1) & & & CS(3$-$2)& \\
\hline
       &$\int T_{\rm MB}{\rm d}v$ & $v_{\rm LSR}$ & FWHM & $\int T_{\rm MB}{\rm d}v$ & $v_{\rm LSR}$ & FWHM \\
       & (K \kms ) & (\kms ) & (\kms ) & (K \kms ) & (\kms ) & (\kms )  \\
\hline
16535$-$4300 &$^{\clubsuit}$ 1.38 & $-$122.83 & 2.0 & 0.90 & $-$123.06 & 1.67 \\
           &$^{\clubsuit}$ 1.02 & $-$88.16 & 2.75 & 0.47 & $-$87.91 & 3.00 \\
16573$-$4214 & 3.44 & $-$23.68 & 2.49 & $\leq 1.1$$^{*}$ & $-$ & $-$ \\
16581$-$4212 & $\leq 0.32$  & $-$ & $-$ & $\leq 0.35$ & $-$ & $-$ \\
17033$-$4035 & 3.22 & $-$114.77 & 1.98 & 1.35 & $-$114.85 & 1.44 \\
17036$-$4033 &$^{\clubsuit}$ 1.26 & $-$80.17 & 3.15 & 0.64 & $-$80.13 & 3.4 \\
           & 1.18 & $-$38.49 & 3.23 & $\leq 0.39$ & $-$ & $-$ \\
17040$-$3959$^{\spadesuit}$ & 2.62 & $-$0.40($-$6,6) & 4.66 & 2.24 & $-$0.21($-$6,6) & 5.50 \\
17082$-$4114 & 9.15 & $-$20.18 & 3.50 & $\leq 0.75$$^{*}$ & $-$ & $-$ \\
17114$-$3804 & $\leq 0.37$ & $-$ & $-$ & $\leq 0.40$ & $-$ & $-$ \\
17140$-$3747 & $\leq 0.36$ & $-$ & $-$ & $\leq 0.38$ & $-$ & $-$ \\
17141$-$3606 & 0.73 & $-$3.97 & 1.6 & 0.19 & $-$3.74 & 0.8 \\
17156$-$3607 & 1.07 & $-$3.15 & 2.4 & 0.68 & $-$3.26 & 2.4 \\
17211$-$3537 & 2.67 & $-$69.58 & 3.37 & 2.00 & $-$69.60 & 4.1 \\
17218$-$3704 & 0.82 & $-$20.68 & 1.6 & 0.60 & $-$20.99 & 2.47 \\
17225$-$3426 &$^{\clubsuit}$ 4.97 & $-$3.82  & 1.91 & 3.51 & $-$4.00 & 1.73 \\
           & 1.30 & $-$0.44  & 3.16 & $\leq 0.50$ & $-$ & $-$ \\
17230$-$3531 & 4.29 & $-$91.89 & 4.2 & 3.02 & $-$91.52 & 3.5 \\
17256$-$3631 & 3.29 & $-$12.45 & 3.13 & 3.29 & $-$12.83 & 3.22 \\
           &$^{\clubsuit}$ 8.02 & $-$8.69 & 2.60 & 7.91 & $-$8.80 & 2.50 \\
17285$-$3346 & 0.71 & $-$16.07 & 2.61 & $\leq 0.38$ & $-$ & $-$ \\
           &$^{\clubsuit}$ 1.84 & 17.60 & 3.9 & 0.76 & 17.15 & 2.62 \\
17338$-$3044 & 1.35 & $-$8.15 & 2.5 & 0.65 & $-$8.04 & 2.36 \\
17355$-$3241 &$^{\clubsuit}$ 2.62 & $-$3.68 & 2.21 & 1.28 & $-$3.82 & 2.15 \\
           & 0.42 &  5.87 & 1.53 & $\leq 0.40$ & $-$ & $-$ \\
17368$-$3057 & $\leq 0.37$ & $-$ & $-$ & $\leq 0.43$ & $-$ & $-$ \\
17377$-$3109 & 10.57($-$8,10) & 0.20 & 8.94 & 10.85 & 0.58 & 8.94 \\
17410$-$3019 & 3.54 & $-$22.15 & 2.16 & 1.89 & $-$22.05 & 1.74 \\
17419$-$3207 & $\leq 0.41$ & $-$ & $-$ & $\leq 0.43$ & $-$ & $-$ \\
17425$-$3017 & 2.34 & $-$20.23 & 2.15 & 2.20 & $-$20.30 & 3.1 \\
\hline
\end{tabular}
\end{center}
$^{\diamondsuit}$ the typical errors from the fits are:
$\sim 0.01-0.07$ K \kms\
in $\int T_{\rm MB}{\rm d}v$, $\sim 0.01 - 0.1$ \kms\ in $v_{\rm LSR}$,
and  $\sim 0.01 - 0.1$ \kms\ in FWHM\\
$^{\spadesuit}$ line parameters derived from moment integrals over the 
velocity range indicated in Cols.~2 and 5 between brackets\\
$^{\clubsuit}$ component observed also in the \CIII\ (1$-$0) and/or (2$-$1)
spectrum \\
$^{*}$ Not observed in the (3$-$2) line. The upper limits refer to 
the (5$-$4) transition \\
\end{table*}
\small
\begin{table*}
\caption[] {\CIII\ line parameters$^{(\diamondsuit)}$. 
Typical rms noise in the spectra is $\sim 0.05$ and $\sim 0.06$ K
for the (1--0) and (2--1) lines, respectively.
N.O. = not observed}
\label{tcoline}
\begin{center}
{\scriptsize
\begin{tabular}{ccccccccccc}
\hline \hline
source & & & \CIII\ (1$-$0) & & & & & \CIII\ (2$-$1) & \\
\hline
     & vel. range & $\int T_{\rm MB}{\rm d}v$ & $v_{\rm LSR}$ & FWHM & $\tau_{10}$ & vel. range & $\int T_{\rm MB}{\rm d}v$ & $v_{\rm LSR}$ & FWHM & $\tau_{21}$ \\
       & (\kms ) & (K \kms ) & (\kms ) & (\kms ) & & (\kms ) & (K \kms ) & (\kms ) & (\kms ) & \\
\hline
08211$-$4158 & 5,14 & 1.08 & 10.99 & 1.7 & $\ll 1$ & 8,14 & 2.71 & 11.04 & 1.62 & $\ll 1$ \\
08247$-$4223 & & $\leq 0.9$ & $-$ & $-$ & $-$ & N.O. & & & & \\
08477$-$4359 & 4,11 & 0.75 & 8.7 & 2.24 & $-$ & N.O. & & & & \\
08488$-$4457 & N.O. &  & & & & & & & & \\
08563$-$4225 & 3,10 & 2.95 & 7.42 & 1.62 & $\ll 1$ & 5,11 & 6.80 & 7.46 & 1.89 & 0.2 \\
09014$-$4736 & & $\leq 0.5$ & $-$ & $-$ & $-$ & N.O. & & & & \\
09026$-$4842 & 1,7 & 0.75 & 4.82 & 0.88 & $-$ & 3,8 & 1.78 & 5.04 & 1.66 & $\ll 1$ \\
09131$-$4723 & 0,7 & 1.39 & 4.13 & 2.17 & $-$ & 1,7 & 2.80 & 4.17 & 1.59 & $\ll 1$ \\
09166$-$4813 & 2,9 & 0.58 & 6.03 & 1.81 & $-$ & 3,9 & 0.73 & 6.19 & 2.8 & $-$ \\
09209$-$5143 & 33,41 & 0.61  & 37.01 & 2.00 & $-$ & & $\leq 0.8$ & $-$ & $-$ & $-$ \\
10088$-$5730 & N.O. & & & & & & & & & \\
10095$-$5843 & $-$9,$-$1 & 0.89 & $-$4.35 & 1.85 & $-$ & $-$7,$-$2 & 1.71 & $-$4.29 & 1.1 & 1.2 \\
10102$-$5706 & & $\leq 0.6$ & $-$ & $-$ & $-$ & N.O. & & & & \\
10123$-$5727 & & $\leq 0.4$ & $-$ & $-$ & $-$ & N.O. & & & & \\
10156$-$5804 & & $\leq 0.5$ & $-$ & $-$ & $-$ & N.O. & & & & \\
10277$-$5730 & & $\leq 0.4$ & $-$ & $-$ & $-$ &  & $\leq 0.6$ & $-$ & $-$ & $-$ \\
10308$-$6122 & & $\leq 0.5$ & $-$ & $-$ & $-$ & N.O. & & $-$ & $-$ & $-$ \\
10317$-$5936 & & $\leq 0.5$ & $-$ & $-$ & $-$ & N.O. & & $-$ & $-$ & $-$ \\
10439$-$5941 & & $\leq 0.5$ & $-$ & $-$ & $-$ & $-$17,$-$12 & 1.56 & $-$14.54 & 1.7 & $-$ \\
10521$-$6031 & & $\leq 0.4$ & $-$ & $-$ & $-$ & & $\leq 0.6$ & $-$ & $-$ & $-$ \\
10537$-$5930 & & $\leq 0.5$ & $-$ & $-$ & $-$ & N.O. &  & $-$ & $-$ & $-$ \\
10545$-$6244 & & $\leq 0.4$ & $-$ & $-$ & $-$ & & $\leq 0.7$ & $-$ & $-$ & $-$ \\
10548$-$5929 & & $\leq 0.5$ & $-$ & $-$ & $-$ & & $\leq 0.8$ & $-$ & $-$ & $-$ \\
10554$-$6237 & $-$20,$-$14 & 0.7 & $-$17.0 & 2.1 & $-$ & $-$20,$-$14 & 1.62 & $-$16.95 & 2.1 & $-$ \\
10555$-$5949 & N.O. & & & & & & & & & \\
10572$-$6018 & & $\leq 0.4$ & $-$ & $-$ & $-$ & & $\leq 0.6$ & $-$ & $-$ & $-$ \\
10575$-$5844 & & $\leq 0.4$ & $-$ & $-$ & $-$ & & $\leq 0.6$ & $-$ & $-$ & $-$ \\
10591$-$5934 & & $\leq 0.5$ & $-$ & $-$ & $-$ & N.O. & $-$ & $-$ & $-$ & $-$ \\
11265$-$6158 & & $\leq 0.5$ & $-$ & $-$ & $-$ & N.O. & & & & \\
11294$-$6257 & & $\leq 0.5$ & $-$ & $-$ & $-$ & N.O. & & & & \\
11380$-$6311 & & $\leq 0.4$ & $-$ & $-$ & $-$ & $-$14,$-$8 & 1.02 & $-$11.43 & 1.78 & $-$ \\
11396$-$6202 & & $\leq 0.5$ & $-$ & $-$ & $-$ & & N.O. & $-$ & $-$ & $-$ \\
11404$-$6215 & & $\leq 0.4$ & $-$ & $-$ & $-$ & 36,43 & 0.80 & 38.52 & 2.1 & $-$ \\
11476$-$6435 & N.O. & & & & & & & & & \\
12102$-$6133 & & $\leq 0.5$ & $-$ & $-$ & $-$ & N.O. & & $-$ & $-$ & $-$ \\
12295$-$6224 & $-$41,$-$35 & 1.04 & $-$36.86 & 1.3 & 0.2 & $-$40,$-$35 & 2.10 & $-$36.79 & 1.3 & 0.5 \\
12377$-$6237 & & $\leq 0.9$ & $-$ & $-$ & $-$ & N.O. & & $-$ & $-$ & $-$ \\
12434$-$6355 & N.O. & & & & & & & & & \\
13023$-$6213 & $-$48,$-$40 & 0.95 & $-$43.3 & 2.7 & $-$ & N.O. & & & & \\
13039$-$6108 & $-$30,$-$24 & 0.40 & $-$26.2 & 0.87 & $-$ & $-$28,$-$23 & 1.20 & $-$26.24 & 0.93 & 0.6 \\
13078$-$6247 & N.O. & & & & & & & & & \\
13106$-$6050 & $-$62,$-$53 & 2.15 & $-$56.87 & 2.11 & $\ll 1$ & $-$60,$-$55 & 3.62 & $-$56.89 & 1.43 & 0.4 \\
13333$-$6234 & $-$16,$-$8 & 1.66 & $-$11.7 & 3.67 & $\ll 1$ & $-$16,$-$7 & 2.51 & $-$11.73 & 3.6 & $\ll 1$ \\
13384$-$6152 & $-$55,$-$47 & 1.73 & $-$50.48 & 1.93 & $\ll 1$ & $-$53,$-$47 & 3.53 & $-$50.27 & 1.65 & 0.3 \\
13395$-$6153 & $-$55,$-$47 & 2.18 & $-$50.44 & 2.60 & 0.3 & $-$54,$-$47 & 5.05 & $-$50.56 & 2.85 & $\ll 1$ \\
13438$-$6203 & $-$56,$-$48 & 0.85 & $-$50.75 & 3.18 & $-$ & $-$53,$-$49 & 0.85 & $-$50.98 & 1.8 & $-$ \\
13481$-$6124 & $-$45,$-$37 & 2.85 & $-$40.7 & 2.6 & 0.3 & $-$44,$-$37 & 5.52 & $-$40.74 & 2.68 & $-$ \\
13558$-$6159 & N.O. & & & $-$ & $-$ & $-$ & $-$ & $-$ & $-$ & $-$ \\
13560$-$6133 & $-$62,$-$55 & 1.19 & $-$57.65 & 2.41 & $-$ & $-$61,$-$56 & 1.92 & $-$58.26 & 1.5 & $-$ \\
14000$-$6104 & $-$63,$-$55 & 0.80 & $-$58.9 & 2.5 & $-$ & $-$63,$-$55& 1.60 & $-$58.2 & 5.3 & $-$ \\
14131$-$6126 & $-$38,$-$33 & 0.69 & $-$34.82 & 2.26 & $-$ & $-$38,$-$33& 0.99 & $-$34.34 & 1.44 & $-$ \\
14166$-$6118 & $-$45,$-$37 & 1.82 & $-$41.23 & 2.53 & $\ll 1$ & $-$43,$-$38 & 2.30 & $-$41.24 & 2.5 & $\ll 1$ \\
\hline
\end{tabular}
}
\end{center}
\end{table*}
\addtocounter{table}{-1}
\begin{table*}
\caption[] {Continued. }
\begin{center}
{\scriptsize
\begin{tabular}{ccccccccccc}
\hline \hline
source & & & \CIII\ (1$-$0) & & & & & \CIII\ (2$-$1) & \\
\hline
     & vel. range  & $\int T_{\rm MB}{\rm d}v$ & $v_{\rm LSR}$ & FWHM & $\tau_{10}$ & vel. range & $\int T_{\rm MB}{\rm d}v$ & $v_{\rm LSR}$ & FWHM & $\tau_{21}$ \\
       & (\kms ) & (K \kms ) & (\kms ) & (\kms ) & & (\kms ) & (K \kms ) & (\kms ) & (\kms ) & \\
\hline
14183$-$6050 & & $\leq 0.4$ & $-$ & $-$ & $-$ & N.O. & & & & \\
14198$-$6115 & N.O. & & & & & & & & & \\
14201$-$6044 & $-$54,$-$47 & 1.16 & $-$49.58 & 1.75 & $-$ & $-$52,$-$46 & 2.64 & $-$49.68 & 1.7 & $\ll 1$ \\
14395$-$5941 & $-$47,$-$39 & 0.77 & $-$42.13 & 2.9 & $-$ & $-$45,$-$40& 1.15 & $-$42.23 & 0.8 & $-$ \\
14412$-$5948 & N.O. & & & & & & & & & \\
14425$-$6023 & $-$50,$-$42 & 1.70 & $-$45.22 & 2.2 & $\ll 1$ & $-$49,$-$43 & 4.00 & $-$45.89 & 2.32 & $-$ \\
14591$-$5843 & & $\leq 0.4$ & $-$ & $-$ & $-$ & & $\leq 0.6$  & $-$ & $-$ & $-$ \\ 
15038$-$5828 & $-$72,$-$65 & 0.65 & $-$68.09 & 2.04 & $-$ & $-$70,$-$65& 1.23 & $-$67.4 & 1.27 & $-$ \\
15072$-$5855 & $-$45,$-$39 & 1.35 & $-$41.64 & 1.6 & $-$ & $-$45,$-$39& 2.68 & $-$41.8 & 1.4 & $-$ \\
15100$-$5903 & $-$55,$-$48 & 0.89 & $-$51.17 & 1.94 & $-$ & $-$54,$-$49& 0.62 & $-$51.26 & 1.58 & $-$ \\
15178$-$5641 & & $\leq 0.4$ & $-$ & $-$ & $-$ & & $\leq 0.5$ & $-$ & $-$ & $-$ \\
15219$-$5658 & & $\leq 0.4$ & $-$ & $-$ & $-$ & $-$20,$-$12& 1.37 & $-$16.0 & 3.0 & $-$ \\
15239$-$5538 & & $\leq 0.4$ & $-$ & $-$ & $-$ & & $\leq 0.6$ & $-$ & $-$ & $-$ \\
15246$-$5612 & $-$70,$-$60 & 2.05 & $-$65.32 & 2.6 & $\ll 1$ & $-$70,$-$60& 3.48 & $-$65.40 & 1.9 & 0.5 \\
15262$-$5541 & $-$58,$-$52 & 0.73 & $-$54.3 & 1.6 & $-$ & $-$56,$-$52& 0.90 & $-$54.17 & 1.5 & $-$ \\
15347$-$5518 & $-$65,$-$58 & 1.51 & $-$61.4 & 2.1 & $-$ & $-$65,$-$58& 2.12 & $-$61.39 & 1.7 & $\ll 1$ \\
15371$-$5458 & & $\leq 0.4$ & $-$ & $-$ & $-$ & & $\leq 0.6$ & $-$ & $-$ & $-$ \\
15470$-$5419 & $-$68,$-$57& 1.88 & $-$61.8 & 5.3 & $-$ & $-$68,$-$57& 1.94 & $-$61.3 & 2.27 & 0.15 \\
15506$-$5325 & N.O. & & & & & & & & &\\
15519$-$5430 & $-$42,$-$33 & 2.57 & $-$36.64 & 2.52 & $\ll 1$ & $-$40,$-$33 & 7.18 & $-$36.71 & 2.24 & $\ll 1$ \\
15557$-$5337 & $-$51,$-$42 & 4.18 & $-$47.0 & 4.0 & $\ll 1$ & $-$52,$-$43 & 14.96 & $-$46.88 & 3.53 & 0.4 \\
15571$-$5218 & & $\leq 0.4$ & $-$ & $-$ & $-$ & & $\leq 0.7$ & $-$ & $-$ & $-$ \\
15579$-$5303 & $-$54,$-$45& 2.29 & $-$49.4 & 4.0 & $\ll 1$ &$-$54,$-$45 & 3.29 & $-$49.12 & 3.6 & $\ll 1$ \\
15583$-$5314 & $-$82,$-$75& 1.52 & $-$77.76 & 1.66 & $\ll 1$ & $-$82,$-$75& 2.88 & $-$77.90 & 1.73 & $\ll 1$ \\
16061$-$5048 & $-$56,49 & 1.80 & $-$51.78 & 1.60 & $\ll 1$ & $-$54,$-$48 & 2.1  & $-$51.80 & 1.91 & $\ll 1$ \\
16082$-$5031 & $-$45,$-$39 & 1.88 & $-$40.96 & 1.62 & 0.3 & $-$44,$-$39& 2.27 & $-$41.01 & 1.67 & 0.3 \\
16093$-$5015 & $-$47,$-$40& 1.47 & $-$42.97 & 1.41 & $\ll 1$ & $-$46,$-$40& 2.44 & $-$43.02 & 1.43 & $\ll 1$ \\
16093$-$5128 & $-$101,$-$94& 1.42 & $-$96.94 & 2.14 & $\ll 1$ & $-$99,$-$93& 1.77 & $-$96.87 & 2.14 & $\ll 1$ \\
16106$-$5048 & $-$93,$-$85& 1.10 & $-$88.76 & 2.00 & $\ll 1$ & $-$92,$-$85& 1.37 & $-$89.05 & 2.17 & $\ll 1$ \\
16107$-$4956 & $-$85,$-$79& 0.59 & $-$82.9 & 2.0  & $-$ & $-$85,$-$79& 0.33 & $-$82.7 & 0.6 & $-$ \\
16148$-$5011 & $-$49,$-$41& 1.06 & $-$44.9 & 2.2 & $-$ & $-$48,$-$41& 3.00 & $-$44.84 & 2.0 & 0.4 \\
16153$-$5016 & & $\leq 0.4$ & $-$ & $-$ & $-$ & $-$44,$-$38& 1.03 & $-$41.43 & 1.2 & 1.2 \\
16170$-$5053 & & $\leq 0.9$ & $-$ & $-$ & $-$ & & $\leq 1.0$ & $-$ & $-$ & $-$ \\
16187$-$4932 & $-$53,$-$43& 1.20 & $-$48.27 & 1.6 & $-$ & & $\leq 0.7$ & $-$ & $-$ & $-$ \\
16194$-$4934 & $-$93,$-$82& 2.02 & $-$85.1 & 3.4 & 1.2 & $-$93,$-$82& 1.44 & $-$84.86 & 2.77 & $\ll 1$ \\
16204$-$4916 & $-$74,$-$66& 1.01 & $-$69.9 & 4.9 & $-$ &$-$74,$-$65 & 1.78 & $-$69.3 & 3.23 & $-$ \\
16204$-$4943 & N.O. & & & & & & & & & \\
16218$-$4931 & $-$45,$-$33& 1.59 & $-$38.2 & 4.8 & 0.7 & $-$45,$-$33& 2.34 & $-$37.6 & 4.03 & $\ll 1$ \\
16219$-$4848 & & $\leq 0.4$ & $-$ & $-$ & $-$ & $-$83,$-$76& 0.75 & $-$79.2 & 0.9 & $-$ \\
16231$-$4819 & & $\leq 0.6$ & $-$ & $-$ & $-$ & & $\leq 1.0$ & $-$ & $-$ & $-$ \\
16232$-$4917 & & $\leq 0.4$ & $-$ & $-$ & $-$ & $-$49,$-$44 & 1.97 & $-$46.38 & 1.63 & $-$ \\
16252$-$4853 & $-$50,$-$42& 0.83 & $-$45.66 & 1.78 & $-$ & $-$50,$-$42& 0.77 & $-$45.3 & 2.07 & $-$ \\
16254$-$4844 & $-$47,$-$38& 1.77 & $-$40.84 & 1.6 & 0.3 & & 2.24 & $-$40.91 & 1.42 & $\ll 1$ \\
16344$-$4605 & $-$67,$-$56& 1.68 & $-$62.1 & 3.8 & $-$ & $-$67,$-$56& 2.34 & $-$61.83 & 2.8 & $-$ \\
16358$-$4614 & & $\leq 0.6$ & $-$ & $-$ & $-$ & & $\leq 1.0$ & $-$ & $-$ & $-$ \\
16363$-$4645 & & $\leq 0.5$ & $-$ & $-$ & $-$ & $-$69,$-$61& 0.87 & $-$65.3 & 3.0 & $-$ \\
16369$-$4810 & $-$45,$-$32& 2.70 & $-$38.54 & 5.0 & $-$ & $-$43,$-$34& 2.27 & $-$38.4 & 3.35 & $-$ \\
16403$-$4614 &$-$124,$-$117 & 0.72 & $-$120.1 & 1.7 & $-$ & $-$124,$-$117& 0.60 & $-$120.1 & 1.34 & $-$ \\
16404$-$4518 & & $\leq 0.6$ & $-$ & $-$ & $-$ & & $\leq 1.0$ & $-$ & $-$ & $-$ \\
16417$-$4445 & & $\leq 0.4$ & $-$ & $-$ & $-$ & $-$59 $-$53& 0.70 & $-$56.49 & 0.66 & $-$ \\
16419$-$4602 & $-$42,$-$35& 1.57 & $-$37.70 & 1.91 & $\ll 1$ &$-$41,$-$34 & 2.35 & $-$37.32 & 1.7 & $\ll 1$ \\
16428$-$4109 & & $\leq 0.5$ & $-$ & $-$ & $-$ & $-$27,$-$24& 0.93 & $-$25.7 & 0.99 & $\ll 1$  \\
16464$-$4359 &$-$83,$-$75 & 1.41 & $-$78.81 & 2.4 & $-$ & $-$83,$-$75& 2.76 & $-$79.4 & 2.05 & 0.7 \\
\hline
\end{tabular}
}
\end{center}
\end{table*}
\addtocounter{table}{-1}
\begin{table*}
\caption[] {Continued. }
\begin{center}
{\scriptsize
\begin{tabular}{ccccccccccc}
\hline \hline
source & & & \CIII\ (1$-$0) & & & & & \CIII\ (2$-$1) & \\
\hline
     & vel. range  & $\int T_{\rm MB}{\rm d}v$ & $v_{\rm LSR}$ & FWHM & $\tau_{10}$ & vel. range & $\int T_{\rm MB}{\rm d}v$ & $v_{\rm LSR}$ & FWHM & $\tau_{21}$ \\
       & (\kms ) & (K \kms ) & (\kms ) & (\kms ) & & (\kms ) & (K \kms ) & (\kms ) & (\kms ) & \\
\hline
16501$-$4314 & $-$123,$-$116& 1.27 & $-$119.0 & 2.1 & $\ll 1$ & $-$123,$-$116& 2.92 & $-$119.0 & 2.3 & 0.3 \\
16535$-$4300 & $-$128,$-$121& 1.69 & $-$122.5 & 1.46 & 0.1 & $-$125,$-$120& 2.1 & $-$122.5 & 1.05 & 0.8 \\ 
           & $-$92,$-$85& 0.68 & $-$88.9 & 4.04 & & $\leq 0.9$ & & & \\
16573$-$4214 & $-$28,$-$20& 2.24 & $-$23.64 & 2.15 & $\ll 1$ & $-$27,$-$20& 4.3 & $-$23.77 & 1.91 & $\ll 1$ \\
16581$-$4212 & N.O. & & & & & & & & & \\
17033$-$4035 &$-$118,$-$112 & 1.66 & $-$114.8 & 1.59 & 0.4 & $-$118,$-$112& 2.83 & $-$114.8 & 1.79 & $\ll 1$ \\
17036$-$4033 & $-$84,$-$78& 0.65 & $-$80.4 & 2.0 & $-$ & $-$84,$-$78& 1.07 & $-$79.85 & 1.42 & $-$  \\
17040$-$3959 & $-$7,4 & 1.22 & 0.02 & 2.5 & $\ll 1$ & $-$5,3& 2.3 & $-$0.42 & 3.23 & $-$ \\
17082$-$4114 & $-$25,$-$15& 2.4 & $-$19.91 & 2.6 & 0.2 & $-$24,$-$16& 3.11 & $-$20.05 & 1.7 & 0.9 \\
17114$-$3804 & & $\leq 0.6$ & $-$ & $-$ & $-$ & & $\leq 0.9$ & $-$ & $-$ & $-$ \\
17140$-$3747 & & $\leq 0.5$ & $-$ & $-$ & $-$ & & $\leq 1.0$ & $-$ & $-$ & $-$ \\
17141$-$3606 & $-$8,$-$1& 0.90 & $-$4.25 & 1.8 & & $\leq 1.0$ & $-$ & $-$ & $-$ & $-$ \\
17156$-$3607 & $-$8,$-$1& 1.1 & $-$3.26 & 1.42 & $-$ & $-$5,$-$1& 1.24 & $-$3.09 & 1.2 & $-$ \\
17211$-$3537 & $-$8,$-$1& 1.09 & $-$69.6 & 2.2 & $-$ & & 1.20 & $-$69.78 & 2.0 & $-$ \\
17218$-$3704 & $-$26,$-$17& 1.15 & $-$20.41 & 1.3 & 0.8 &$-$23,$-$19 & 1.76 & $-$20.6 & 1.0 & 0.6 \\
17225$-$3426 & $-$7,0& 0.64 & $-$3.42 & 1.6 & $-$ & $-$7,0& 2.90 & $-$3.6 & 2.7 & $\ll 1$ \\
17230$-$3531 & $-$96,$-$88& 2.68 & $-$91.6 & 2.8 & $\ll 1$ & $-$96,$-$88& 4.12 & $-$91.62 & 1.57 & 0.6 \\
17256$-$3631 & $-$12,$-$5& 0.77 & $-$8.11 & 1.3 & $-$ & & 2.86 & $-$8.46 & 1.72 & $-$ \\
17285$-$3346 & & $\leq 0.4$ & $-$ & $-$ & $-$ & 15,21& 1.22 & 18.05 & 1.74 & $-$ \\
17338$-$3044 & & $\leq 0.5$ & $-$ & $-$ & $-$ & $-$10,$-$5& 0.75 & $-$7.9 & 1.6 & $-$  \\
17355$-$3241 & $-$9,0& 1.59 & $-$3.97 & 1.5 & $\ll 1$ & $-$8,$-$2& 1.91 & $-$3.98 & 1.6 & $\ll 1$ \\
17368$-$3057 & & $\leq 0.6$ & $-$ & $-$ & $-$ & & $\leq 0.9$ & $-$ & $-$ & $-$ \\
17377$-$3109 &$-$5,8 & 4.01 & 2.74 & 5.7 & $\ll 1$ & $-$5,10& 5.83 & 1.21 & 7.1 & $\ll 1$ \\
17410$-$3019 &$-$26,$-$19 & 1.63 & $-$22.12 & 1.8 & $\ll 1$ & $-$24,$-$20& 2.92 & $-$22.1 & 1.23 & 0.5 \\
17419$-$3207 & & $\leq 0.6$  & $-$ & $-$ & $-$ & & $\leq 1.0$ & $-$ & $-$ & $-$ \\
17425$-$3017 & $-$24,$-$17& 1.25 & $-$19.93 & 1.4 & $-$ & $-$22,$-$18& 2.2 & $-$20.1 & 1.14 & 0.6 \\
\hline
\end{tabular}
}
\end{center}
$^{\diamondsuit}$ the errors from the fits are:
$\sim 0.02-0.06$ K \kms\
in $\int T_{\rm MB}{\rm d}v$, $\sim 0.01 - 0.1$ \kms\ in $v_{\rm LSR}$,
and  $\sim 0.01 - 0.1$ \kms\ in FWHM\\
\end{table*}

\begin{table*}
\caption[] {``Distance-independent'' parameters of the clumps. The angular diameters,
$\theta$, and integrated flux densities, $F_{\nu}$, have been derived from the 
1.2~mm continuum maps; the rotation temperatures, $T_{\rm rot}$, the \CIII\ column 
densities, $N_{\rm C^{17}O}$, and the H$_{2}$ total column densities, 
$N_{\rm H_{2}}$, have been derived from ${C^{17}O}$ line ratios 
(assuming a \CIII\ mean abundance of 3.9$\times10^{-8}$); the $H_{2}$ volume 
densities, $n_{\rm H_2}$, have been obtained from CS line ratios.}
\label{tdist_ind}
\begin{center}
\begin{tabular}{ccccccc}
\hline \hline
source & $\theta$ & $F_{\nu}$ & $T_{\rm k}$ & $N_{\rm C^{17}O}$ & $N_{\rm H_2}$ & 
$n_{\rm H_2}$ \\
       &  (\asec ) & (Jy) & (K) & ($\times10^{15}$\cmq ) & (\cmq ) & (\cmc ) \\
\hline
08211$-$4158 & 34.3 & 2.26 & 12 & 2.3 & 5.9$\times10^{22}$ & 1.2$\times10^{5}$ \\
08563$-$4225 & 29.5 & 4.09 & 10 & 7.6 & 1.9$\times10^{23}$ & 1.7$\times10^{5}$ \\
09014$-$4736 & 42.1 & 0.79 &      &           &          &  \\
09026$-$4842 & 65.8 & 1.25 & 15 & 1.0 & 2.5$\times10^{22}$ & 6.3$\times10^{5}$ \\
09131$-$4723 & 71.6 & 4.5  & 14 & 1.7 & 4.3$\times10^{22}$ & 4.2$\times10^{5}$ \\
09166$-$4813 &    &  & 11 & 0.4 & 1.1$10\times^{22}$ &  \\
09209$-$5143 & 64.5 & 1.43 &     &          &           & \\
10095$-$5843 & 33.9 & 1.16 & 9 & 1.8 & 4.7$\times10^{22}$ & 1.5$\times10^{5}$ \\
10123$-$5727 & 43.2 & 2.41 &    &       &    & 1.7$\times10^{5}$ \\
10277$-$5730 & 27.0 & 0.63 &     &           &           & 4.9$\times10^{4}$ \\
10308$-$6122 & 33.8 & 0.45 &     &           &           & 2.1$\times10^{5}$ \\
10439$-$5941 & 42.8 & 6.90 &     &           &           & \\
10521$-$6031 & 6.5  & 0.46 &     &           &           & \\
10548$-$5929 & 65.5 & 0.62 &     &           &           & 5.2$\times10^{4}$ \\
10554$-$6237 & 21.5 & 0.73 & 8 & 2.9 & 7.4$\times10^{22}$ & 4.0$\times10^{4}$ \\
10572$-$6018 & 37.5 & 1.31 &     &           &           & 4.2$\times10^{4}$ \\
11265$-$6158 & 30.3 & 0.50 &     &           &           & 2.1$\times10^{5}$ \\
11294$-$6257 & 17.9 & 0.27 &     &           &           & \\
11380$-$6311 & 51.1 & 2.53 &     &           &           & 1.8$\times10^{5}$ \\
11404$-$6215 & 9.5  & 0.70 &     &           &           & \\
12295$-$6224 & 45.9 & 2.4  & 11 & 1.6 & 4.2$\times10^{22}$ & 1.4$\times10^{5}$ \\
12377$-$6237 & 15.1 & 0.53 &     &           &           & \\
13023$-$6213 & 28.5 & 1.39 &     &           &           & \\
13039$-$6108 & 50.9 & 1.38 & 18 & 0.7 & 1.9$\times10^{22}$ & 1.7$\times10^{5}$  \\
13106$-$6050 & 33.5 & 1.95 & 8 & 4.5 & 1.1$\times10^{23}$ & 5.7$\times10^{4}$  \\
13333$-$6234 & 31.8 & 1.5  & 8 & 3.7 & 9.4$\times10^{22}$ & 2.2$\times10^{5}$ \\
13384$-$6152 & 30.0 & 0.45 & 9 & 4.2 & 1.1$\times10^{23}$ & \\
13395$-$6153 & 24.5 & 4.2  & 10 & 6.8 & 1.7$\times10^{23}$ & 8.9$\times10^{4}$ \\
13438$-$6203 & 33.4 & 3.9  & 6 & 1.8 & 4.7$\times10^{22}$ & \\
13481$-$6124 & 25.5 & 7.12 & 18 & 2.8 & 7.1$\times10^{22}$ & 1.2$\times10^{5}$ \\
13560$-$6133 & 18.5 & 3.1  & 7 & 6.0 & 1.6$\times10^{23}$ & 1.5$\times10^{5}$ \\
14000$-$6104 & 50.5 & 4.4  & 11 & 1.3 & 3.3$\times10^{22}$ & 4.2$\times10^{5}$ \\
14131$-$6126 & $\ll 21$ ($^{\#}$)  & 0.27  & 6 & 43.0 & 1.1$\times10^{24}$ & 1.7$\times10^{5}$ \\
14166$-$6118 & 48.9 & 1.06 & 8 & 2.4 & 6.1$\times10^{22}$ & 7.1$\times10^{4}$ \\
14183$-$6050 & 38.4 & 0.35 &     &           &           & 1.7$\times10^{5}$ \\
\hline
\end{tabular}
\end{center}
\end{table*}
\addtocounter{table}{-1}
\begin{table*}
\caption[] {Continued. }
\begin{center}
\begin{tabular}{ccccccc}
\hline
source & $\theta$& $F_{\nu}$ & $T_{\rm rot}$ & $N_{\rm C^{17}O}$ & $N_{\rm H_2}$ & 
$n_{\rm H_2}$ \\
       &  (\asec ) & (Jy) & (K) & ($\times10^{15}$\cmq ) & (\cmq ) & (\cmc ) \\
\hline 
14201$-$6044 &   &  & 23 & 1.3 & 3.4$\times10^{22}$ &  \\
14395$-$5941 & $\ll 21$ ($^{\#}$)  & 0.46 & 6 & 46.1 & 1.2$\times10^{24}$ & 2.8$\times10^{4}$ \\
14425$-$6023 & 20.7 & 1.46 & 8 & 7.4 & 1.9$\times10^{23}$ & \\
15038$-$5828 &   &  & 17 & 0.6 & 1.6$\times10^{22}$ & \\
15072$-$5855 & 23.6 & 1.02 & 8 & 4.8 & 1.2$\times10^{23}$ & 1.2$\times10^{5}$ \\
15100$-$5903 & 28.  & 0.95 & 5 & 2.4 & 6.2$\times10^{22}$ & 2.8$\times10^{4}$ \\
15219$-$5658 & 15.2 & 0.52 &   &           &           & 5.3$10^{4}$ \\
15246$-$5612 & 20.2 & 1.26 & 7 & 8.4 & 2.2$\times10^{23}$ & 4.5$\times10^{4}$ \\
15262$-$5541 & 46.9 & 1.82 & 10 & 0.5 & 1.3$\times10^{22}$ & \\
15347$-$5518 & 48.4 & 3.12 & 8 & 2.1 & 5.3$\times10^{22}$ & 2.2$\times10^{5}$ \\
15371$-$5458 & 43.3 & 0.67 &     &           &          & 3.5$\times10^{5}$ \\
15470$-$5419 & 39.5 & 2.41 & 7 & 3.1 & 8.0$\times10^{22}$ & 2.0$\times10^{5}$ \\
15519$-$5430 & 32.1 & 6.38 & 12 & 6.4 & 1.6$\times10^{23}$ & 2.8$\times10^{5}$ \\
15557$-$5337 & 38.9 & 16.7 & 18 & 10.3 & 2.6$\times10^{23}$ & 3.7$\times10^{4}$ \\
15579$-$5303 & 21.3 & 6.26 & 6 & 9.2 & 2.4$\times10^{23}$ & \\
15583$-$5314 & 39.3 & 0.53 & 10 & 2.7 & 6.9$\times10^{22}$ & 1.3$\times10^{5}$ \\
16061$-$5048 & 27.7 & 1.64 & 6 & 4.8 & 1.2$\times10^{23}$ & \\
16082$-$5031 & 35.1 & 0.54 & 7 & 3.6 & 9.2$\times10^{22}$ & \\
16093$-$5015 & 34.8 & 1.81 & 8 & 2.9 & 7.5$\times10^{22}$ & 8.4$\times10^{4}$ \\
16093$-$5128 & 43.0 & 1.28 & 8 & 2.1 & 5.5$\times10^{22}$ & \\
16106$-$5048 &  &  & 11 & 0.8 & 2.0$\times10^{22}$ &  \\
16107$-$4956 &  &  & 6 &  0.4 & 1.1$\times10^{22}$ &  \\
16148$-$5011 & 12.4 & 2.32 & 8 & 11.6 & 3.0$\times10^{23}$ & \\
16187$-$4932 & 35.2 & 0.66 &     &           &           & \\
16194$-$4934 &  &  & 7 & 1.3 & 3.3$\times10^{22}$ &  \\
16204$-$4916 & 32.0 & 1.37 & 8 & 2.2 & 5.7$\times10^{22}$ & 2.4$\times10^{4}$ \\
16218$-$4931 & 16.4 & 0.69 & 6 & 10.1 & 2.6$\times10^{23}$ & 2.6$\times10^{4}$ \\
16232$-$4917 & 16.6 & 0.45 &     &           &           & 4.5$\times10^{4}$ \\
16252$-$4853 & 43.1 & 1.22 & 6 & 1.2 & 3.2$\times10^{22}$ & 2.4$\times10^{4}$ \\
16254$-$4844 &  &  & 12 & 1.1 & 2.9$\times10^{22}$ &  \\
16344$-$4605 & 26.6 & 2.85 & 7 & 4.7 & 1.2$\times10^{23}$ & \\
16363$-$4645 & 13.0 & 0.22 &     &     &        & 5.7$\times10^{4}$ \\
16369$-$4810 &  & & 8 & 1.8 & 4.7$\times10^{22}$ &  \\
16403$-$4614 &  & & 8 & 0.5 & 1.3$\times10^{22}$ & \\
16419$-$4602 & 17. & 2.27 & 9 & 2.1 & 5.3$\times10^{22}$ & \\
16464$-$4359 & 42. & 1.4  & 10 & 2.3 & 5.9$\times10^{22}$ & 7.1$\times10^{4}$ \\
16501$-$4314 & 70.9 & 2.15 & 15 & 1.7 & 4.3$\times10^{22}$ & 1.0$\times10^{5}$ \\
16535$-$4300 & 42.0 & 0.87 & 8 & 2.5 & 6.5$\times10^{22}$ & 3.5$\times10^{4}$ \\
17033$-$4035 &  &  & 15 & 1.5 & 3.7$\times10^{22}$ &  \\
\hline
\end{tabular}
\end{center}
\end{table*}
\addtocounter{table}{-1}
\begin{table*}
\caption[] {Continued. }
\begin{center}
\begin{tabular}{ccccccc}
\hline
source & $\theta$& $F_{\nu}$ & $T_{\rm rot}$ & $N_{\rm C^{17}O}$ & $N_{\rm H_2}$ & 
$n_{\rm H_2}$ \\
       &  (\asec ) & (Jy) & (K) & ($\times10^{15}$\cmq ) & (\cmq ) & (\cmc ) \\
\hline 
17036$-$4033 &  &  & 10 & 0.7 & 1.9$\times10^{22}$ &  \\
17040$-$3959 & 15.6 & 0.95 & 7 & 8.5 & 2.2$\times10^{23}$ & 2.4$\times10^{4}$ \\
17082$-$4114 & 50. & 7.13 & 8 & 3.1 & 7.9$\times10^{22}$ & \\
17156$-$3607 &  &  & 10 & 0.8 & 2.0$\times10^{22}$ &  \\
17211$-$3537 & 36.8 & 1.1 & 7 & 1.6 & 4.2$\times10^{22}$ & 5.5$\times10^{4}$ \\
17218$-$3704 & 45.4 & 1.44 & 9 & 1.7 & 4.3$\times10^{22}$ & 1.7$\times10^{5}$ \\
17225$-$3426 &  &  & 10 & 1.7 & 4.3$\times10^{22}$ &  \\
17230$-$3531 & 21.6 & 1.19 & 7 & 10.9 & 2.8$\times10^{23}$ & 4.2$\times10^{4}$ \\
17256$-$3631 & 18.7 & 5.6 & 45 & 2.0 & 5.2$\times10^{22}$ & 7.1$\times10^{4}$ \\
17285$-$3346 & 11.9 & 0.26 &      &           &           & \\
17338$-$3044 & 19.0 & 0.19 &      &           &           & \\
17355$-$3241 & 45.5 & 0.6 & 8 & 2.2 & 5.7$\times10^{22}$ & 4.2$\times10^{4}$ \\
17377$-$3109 & 17.5 & 2.0 & 6 & 22.7 & 5.812$\times10^{23}$ & 1.1$\times10^{5}$\\
17410$-$3019 & 18.7 & 0.16 & 7 & 8.2 & 2.1$\times10^{23}$ & \\
17425$-$3017 & 21.1 & 1.59 & 7 & 5.1 & 1.3$\times10^{23}$ & 2.7$\times10^{4}$ \\
\hline
\end{tabular}
\end{center}
$^{\#}$ unresolved sources \\
\end{table*}
\begin{table*}
\caption[] {Distance ($d$), linear size ($D$), luminosity ($L$), and
dust temperature ($T_{\rm d}$) of all
sources detected in CS. A ``$-$'' in the columns of $D$ and $T_{\rm d}$ 
indicates that we could not derive any source angular diameter.}
\label{tdist_dep}
\begin{center}
\begin{tabular}{cccccccc}
\hline \hline
source & \multicolumn{2}{c}{$d$} & \multicolumn{2}{c}{$D$} & \multicolumn{2}{c}{$L$} & $T_{\rm d}$ \\
       & \multicolumn{2}{c}{(kpc)} & \multicolumn{2}{c}{(pc)} & \multicolumn{2}{c}{($\times 10^{3}L_{\odot}$)} & (K) \\
       & near  &  far  & near  & far  & near  & far &  \\
\hline 
08211$-$4158  & & 1.72 & & 0.29 & & 3.01 & 28 \\ 
08247$-$4223  & & 1.40 & & $-$ & & 1.52 & $-$\\
08477$-$4359  & & 1.78 & & $-$ & & 3.33 & $-$\\
08563$-$4225  & & 1.65 & & 0.24 & & 3.16 & 32 \\
09014$-$4736  & & 1.32 & & 0.27 & & 3.55 & 37 \\
09026$-$4842  & & 1.85 & & 0.59 & & 2.25 & 26 \\

09131$-$4723  & & 1.66 & & 0.58 & & 2.61 & 23 \\
09166$-$4813  & & 2.25 & & $-$ & & 2.11 & $-$ \\
09209$-$5143  & & 6.38 & & 2.0 & & 13.61 & 24 \\
10095$-$5843  & 1.06 & 2.92 & 0.17 & 0.48 & 6.97 & 52.87 & 24 \\
10102$-$5706  & 0.84 & 2.9 & $-$ & $-$ & 0.36 & 4.30 & $-$ \\ 
10123$-$5727  & 0.91 & 2.95 & 0.19 & 0.62 & 2.47 & 26.00 & 34 \\
10277$-$5730  & & 5.78 & & 0.76 & & 3.24 & 33 \\
10308$-$6122($^{*}$) & & 1.18 & & 0.19 & & 1.45 & 35 \\
10317$-$5936  & & 8.88 & & $-$ & & 57.48 & $-$ \\
10439$-$5941  & & 2.60 & & 0.54 & & 40.92 & $-$ \\
10521$-$6031  & & 8.13 & & 0.26 & & 40.26 & 29 \\
10537$-$5930  & & 7.22 & & $-$ & & 35.77 & $-$\\
10545$-$6244($^{*}$)  & & 2.00 & & $-$ & & 3.19 & $-$ \\
10548$-$5929  & & 7.60 & & 0.94 & & 39.42 & 31 \\
10554$-$6237  & & 2.96 & & 0.31 & & 3.19 & 27 \\
10555$-$5949  & & 6.15  & & $-$ & & 16.28 & $-$ \\
10572$-$6018  & & 7.17 & & 1.30 & & 46.65 & 27 \\
10591$-$5934  & & 2.84 & & $-$ & & 19.00 & $-$ \\
11265$-$6158  & & 3.39 & & 0.50 & & 5.07 & 27 \\
11294$-$6257  & & 3.47 & & 0.30 & & 6.33 & 30 \\
11380$-$6311($^{*}$)  & & 1.34 & & 0.33 & & 1.33 & 24 \\
11396$-$6202  & & 11.16 & & $-$ & & 48.57 & $-$ \\
11404$-$6215  & & 10.97 & & 0.51 & & 74.78 & 30 \\
12102$-$6133  & & 4.04 & & $-$ & & 7.45 & $-$ \\
12295$-$6224  & & 4.35 & & 0.97 & & 8.51 & 22 \\
12377$-$6237  & & 10.91 & & 0.79 & & 44.87 & 28 \\
13023$-$6213  & & 4.82 & & 0.67 & & 18.25 & 26 \\
13039$-$6108($^{*}$)  & & 2.44 & & 0.60 & & 2.12 & 26 \\
13106$-$6050  & & 4.95 & & 0.80 & & 48.76 & 33 \\
13333$-$6234  & 0.94 & 9.54 & 0.15 & 1.47 & 2.75 & 284.1 & 35 \\
13384$-$6152  & & 5.33 & & 0.78 & & 11.24 & 26 \\
\hline
\end{tabular}
\end{center}
\end{table*}
\addtocounter{table}{-1}
\begin{table*}
\caption[] {Continued.}
\begin{center}
\begin{tabular}{cccccccc}
\hline \hline
source & \multicolumn{2}{c}{$d$} & \multicolumn{2}{c}{$D$} & \multicolumn{2}{c}{$L$} & $T_{\rm d}$ \\
       & \multicolumn{2}{c}{(kpc)} & \multicolumn{2}{c}{(pc)} & \multicolumn{2}{c}{($\times 10^{3}L_{\odot}$)} & (K) \\
       & near  &  far  & near  & far  & near  & far &  \\
\hline 
13395$-$6153  & & 5.34 & & 0.63 & & 277.2 & 35 \\
13438$-$6203  & & 5.39 & & 0.87 & & 30.46 & 24 \\
13481$-$6124  & 3.62 & 7.31 & 0.45 & 0.91 & 73.15 & 298.4 & 36 \\
13560$-$6133  & & 5.56 & & 0.50 & & 33.44 & 23 \\
14000$-$6104  & & 5.63 & & 1.38 & & 89.36 & 33 \\
14131$-$6126  & 2.71 & 8.86 & $-$ & $-$ & 5.05 & 54.04 & 30 \\
14166$-$6118  & 3.25 & 8.42 & 0.77 & 2.00 & 10.02 & 67.27 & 32 \\
14183$-$6050  & & 3.36 & & 0.63 & & 12.94 & 37 \\
14201$-$6044  & 4.05 & 7.75 & $-$ & $-$ & 7.56 & 27.67 & $-$ \\
14395$-$5941  & 3.16 & 9.18 & $-$ & $-$ & 5.20 & 43.90 & 30 \\
14425$-$6023  & 3.43 & 8.92 & 0.34 & 0.90 & 7.30 & 49.38 & 26 \\
14591$-$5843  & 2.19 & 10.68 & $-$ & $-$ & 3.66 & 87.14 & $-$ \\
15038$-$5828  & 5.00 & 8.00 & $-$ & $-$ & 11.3 & 29.0 & $-$ \\
15072$-$5855($^{*}$)  & & 3.00 & & 0.34 & & 4.97 & 30 \\
15100$-$5903($^{*}$)  & & 3.65 & & 0.50 & & 11.92 & 33 \\
15178$-$5641  & 2.09 & 11.37 & $-$ & $-$ & 5.80 & 161.8 & $-$ \\
15219$-$5658  & 1.23 & 12.29 & 0.09 & 0.91 & 0.90 & 89.55 & 28 \\
15239$-$5538  & 3.27 & 10.43 & 0.41 & 1.29 & 13.49 & 137.2 & 28 \\
15246$-$5612  & 4.48 & 9.17 & 0.44 & 0.90 & 37.82 & 158.5 & 32 \\
15262$-$5541  & & 3.72 & & $-$ &  & 5.50 & 22 \\
15347$-$5518  & 4.16 & 9.79 & 0.98 & 2.30 & 25.68 & 142.3 & 35 \\
15371$-$5458  & & $-$($^{\dagger}$) &  & $-$ &  & $-$ & $-$ \\
15470$-$5419  & 4.12 & 10.16 & 0.79 & 1.95 & 27.90 & 169.7 & 32 \\
15519$-$5430($^{*}$)  & & 2.65 & & 0.41 & & 18.62 & 30 \\
15557$-$5337  & 3.29 & 11.21 & 0.62 & 2.11 & 345.0 & 4005.3 & 40 \\
15571$-$5218  & 6.63 & 8.03 & $-$ & $-$ & 22.55 & 33.08 & $-$ \\
15579$-$5303  & 3.48 & 11.12 & 0.36 & 1.15 & 19.89 & 203.1 & 30 \\
15583$-$5314  & 4.99 & 9.60 & 0.95 & 1.83 & 12.76 & 47.23 & 30 \\
16061$-$5048  & 3.61 & 11.35 & 0.49 & 1.53 & 7.92 & 78.34 & 28 \\
16082$-$5031  & 3.02 & 12.0 & 0.51 & 2.04 & 21.34 & 336.9 & 37 \\
16093$-$5015  & 3.14 & 11.92 & 0.53 & 2.01 & 7.20 & 103.8 & 25 \\
16093$-$5128  & 6.05 & 8.90 & 1.26 & 1.86 & 28.92 & 62.58 & 30 \\
16106$-$5048  & 5.52 & 9.51 & $-$ & $-$ & 13.26 & 39.36 & $-$ \\
16107$-$4956  & 5.21 & 9.90 & $-$ & $-$ & 16.24 & 58.64 & $-$ \\
16148$-$5011  & 3.27 & 11.89 & 0.20 & 0.72 & 44.17 & 584.0 & 38 \\
16153$-$5016  & 3.07 & 12.08 & $-$ & $-$ & 33.14 & 513.1 & $-$ \\
16170$-$5053  & 3.75 & 11.37 & $-$ & $-$ & 27.49 & 252.7 & $-$ \\
16187$-$4932  & 4.13 & 11.15 & 0.71 & 1.90 & 11.68 & 85.13 & 31 \\
\hline
\end{tabular}
\end{center}
\end{table*}
\addtocounter{table}{-1}
\begin{table*}
\caption[] {Continued. }
\begin{center}
\begin{tabular}{cccccccc}
\hline \hline
source & \multicolumn{2}{c}{$d$} & \multicolumn{2}{c}{$D$} & \multicolumn{2}{c}{$L$} & $T_{\rm d}$ \\
       & \multicolumn{2}{c}{(kpc)} & \multicolumn{2}{c}{(pc)} & \multicolumn{2}{c}{($\times 10^{3}L_{\odot}$)} & (K) \\
       & near  &  far  & near  & far  & near  & far &  \\
\hline 
16194$-$4934  & 5.30 & 9.98 & $-$ & $-$ & 14.98 & 53.11 & $-$ \\
16204$-$4916  & 4.56 & 10.76 & 0.71 & 1.67 & 25.04 & 139.4 & 30 \\
16218$-$4931  & 2.90 & 12.42 & 0.23 & 0.99 & 4.77 & 87.53 & 30 \\
16219$-$4848  & 5.06 & 10.33 & $-$ & $-$ & 9.11 & 37.97 & $-$ \\
16231$-$4819  & 3.88 & 11.52 &  $-$ & $-$ & 12.87 & 113.4  & $-$ \\
16232$-$4917  & 3.42 & 11.94 & 0.28 & 0.96 & 13.08 & 159.4 & 34 \\
16252$-$4853  & 3.41 & 12.02 & 0.71 & 2.51 & 17.27 & 214.6 & 33 \\
16254$-$4844  & 3.40 & 12.04 & $-$ & $-$ & 7.90 & 99.12 & $-$ \\
16344$-$4605  & 4.36 & 11.44 & 0.56 & 1.48 & 11.74 & 80.83 & 24 \\
16363$-$4645  & 4.51 & 11.25 & $-$ & $-$ & 84.79 & 527.6 & 42 \\ 
16369$-$4810($^{*}$)  & & 3.12 &  & $-$ & & 5.107 & $-$ \\
16403$-$4614  & 6.28 & 9.03  & $-$ & $-$ & 36.13 & 74.7 & $-$ \\
16417$-$4445  & 4.22 & 11.77 & $-$ & $-$ & 14.39 & 111.9 & $-$ \\
16419$-$4602  & 3.15 & 12.74 & 0.78 & 3.15 & 17.90 & 292.8 & 30 \\ 
16428$-$4109($^{*}$)  & & 2.67 & $-$ & $-$ & & 3.97 & $-$ \\
16464$-$4359  & 5.27 & 10.82 & 1.07 & 2.20 & 21.21 & 89.41 & 25 \\
16501$-$4314  & 6.70 & 9.49 & 2.30 & 3.26 & 102.8 & 206.2 & 30 \\
16535$-$4300  & 6.84 & 9.41 & 1.39 & 1.92 & 25.99 & 49.19 & 25 \\
16573$-$4214  & 2.60 & 13.73 & $-$ & $-$ & 21.98 & 613.0 & $-$ \\
17033$-$4035  & 6.69 & 9.79 & $-$ & $-$ & 62.26 & 133.3 & $-$ \\
17036$-$4033  & 5.70 & 10.79 & $-$ & $-$ & 25.15 & 90.12 & $-$ \\
17040$-$3959($^{*}$)  &  & 16.48 & & 1.25 & & 988.0 & 23 \\
17082$-$4114($^{*}$)  & & 2.51 & & 0.62 & & 8.57 & 24\\
17141$-$3606($^{*}$)  & & 0.90 &  & $-$ & & 0.75 & $-$ \\
17156$-$3607($^{*}$)  & & 0.74 &  & $-$ & & 0.92 & $-$ \\
17211$-$3537  & 6.23 & 10.61 & 1.11 & 1.89 & 32.90 & 95.42 & 30  \\
17218$-$3704($^{*}$)  & & 3.32 & & 0.73 & & 16.63 & 33 \\
17225$-$3426  & 1.1 & 15.77 & $-$ & $-$ & 14.05 & 2887.0 & $-$ \\
17230$-$3531  &  & $-$($^{\dagger}$) &  & $-$ & & $-$ &$-$ \\
17256$-$3631($^{*}$)  & & 1.96 & & 0.18 & & 63.77 & 40 \\
17285$-$3346  & 3.74 & 13.17 & 0.22 & 0.76 & 5.86 & 72.71 & 30 \\
17338$-$3044  & 3.99 & 13.00 & 0.37 & 1.20 & 9.24 & 98.08 & 30 \\
17355$-$3241($^{*}$)  & & 1.71 & & 0.38 & & 2.30 & 31 \\
17377$-$3109  & 0.84 & 16.15 & 0.07 & 1.37 & 1.51 & 558.2 & 31 \\
17410$-$3019  & & $-$($^{\dagger}$) & & $-$ & & $-$ & 38 \\
17425$-$3017  & & $-$($^{\dagger}$) & & $-$ & & $-$ & 25 \\
\hline
\end{tabular}
\end{center}
($^{*}$) resolved distance ambiguity \\
($^{\dagger}$) no distance estimate \\ 
\end{table*}

\begin{table*}
\caption[] {Clumps masses estimated from 1.2~mm continuum ($M_{\rm cont}$), 
assuming virial equilibrium ($M_{\rm vir}$), from \CIII\ ($M_{\rm C^{17}O}$) 
and from CS ($M_{\rm CS}$). All values are in $M_{\odot}$ units.}
\label{tmass}
\begin{center}
\begin{tabular}{ccccccccc}
\hline \hline
source & \multicolumn{2}{c}{$M_{\rm cont}$} & \multicolumn{2}{c}{$M_{\rm vir}$} & \multicolumn{2}{c}{$M_{\rm C^{17}O}$} & \multicolumn{2}{c}{$M_{\rm CS}$} \\
       & near  &  far  & near  & far  & near  & far  & near & far \\
\hline 
08211$-$4158 &    & 420 &    & 93 &    & 37 &    & 66 \\
08563$-$4225 &    & 596 &   & 354 &   & 84 &    & 55 \\
09014$-$4736 &    & 62 &  & 127 &  & $-$ &  & $-$ \\
09026$-$4842 &    & 294 &  & 208 &  & 69 &  & 497 \\
09131$-$4723 &    & 993 &  & 357 &  & 110 &  & 190 \\
09209$-$5143 &    & 4419 &    & 735 &   & $-$ &   & $-$ \\
10095$-$5843 & 99 & 1046 & 105 & 290 & 11 & 83 & 18 & 394 \\
10123$-$5727 & 100 & 1044 & 161 & 523 & $-$ & $-$ & 26 & 917 \\
10277$-$5730 &    & 1087 &    & 497 &   & $-$ &   & 495 \\
10308$-$6122 &    & 30 &    & 109 &   & $-$ &    & 34 \\
10439$-$5941 &    & 2697 &    & 332 &    & $-$ &    & 362 \\
10521$-$6031 &    & 1831 &    & 142 &    & $-$ &    & $-$ \\
10548$-$5929 &    & 1991 &    & 719 &    & $-$ &    & 993 \\
10554$-$6237 &  & 420 &  & 154 &  & 55 &    & 27 \\
10572$-$6018 &  & 4422 &  & 925 &  & $-$ &  & 2144 \\
11265$-$6158 &  & 377 &  & 201 &  & $-$ &  & 152 \\
11294$-$6257 &  & 188 &  & 126 &  & $-$ &  & $-$ \\
11380$-$6311 &  & 345 &  & 256 &  & $-$ &  & 148 \\
11404$-$6215 &  & 4871 &  & 447 &  & $-$ &  & $-$ \\
12295$-$6224 &  & 3847 &  & 496 &  & 304 &  & 913 \\
13023$-$6213 &  & 2220 &  & 556 &  & $-$ &  & 692 \\
13039$-$6108 &  & 565 &  & 183 &  & 53 &  & 86 \\
13106$-$6050 &  & 2467 &  & 474 &  & 577 &  & 666 \\
13333$-$6234 & 64 & 6581 & 314 & 3183 & 15 & 1575 & 17 & 1634 \\
13384$-$6152 &  & 879 &  & 538 &  & 511 &  & 372 \\
13395$-$6153 &  & 5773 &  & 977 &  & 547 &  & 514 \\
13438$-$6203 &  & 8602 &  & 2652 &  & 278 &  & $-$ \\
13481$-$6124 & 4353 & 17748 & 460 & 929 & 410 & 4400 & 248 & 2050 \\
13560$-$6133 &  & 7674 &  & 859 &  & 300 &  & 417 \\
14000$-$6104 &  & 7202 &  & 2676 &  & 484 &  & 2565 \\
14131$-$6126 & 115 & 1226 & $-$ & $-$ & 37 & 399 & $-$ & $-$ \\
14166$-$6118 & 600 & 4024 & 709 & 1836 & 281 & 1888 & 743 & 13010 \\
14183$-$6050 &  & 179 &  & 79 &  & $-$ &  & 97 \\
14395$-$5941 & 266 & 2241 & $-$ & $-$ & 54 & 457 & $-$ & $-$ \\
14425$-$6023 & 1181 & 7988 & 354 & 921 & 176 & 1189 & 63 & 1174 \\
15072$-$5855 &  & 531 &  & 173 &  & 113 &  & 107 \\
15100$-$5903 &  & 654 &  & 297 &  & 118 &  & 162 \\
15219$-$5658 & 49 & 4934 & 152 & 1521 & $-$ & $-$ & $-$ & $-$ \\
15246$-$5612 & 1354 & 5674 & 431 & 883 & 323 & 1352 & 88 & 750 \\
15262$-$5541 &  & 2134 & & 350 & & $-$ & & $-$ \\
15347$-$5518 & 2603 & 14414 & 947 & 2229 & 395 & 2186 & 484 & 6257 \\
15470$-$5419 & 2191 & 13322 & 1989 & 4905 & 387 & 2356 & 227 & 3411 \\
15519$-$5430 &  & 2591 & & 357 & & 218 & & 448 \\
15557$-$5337 & 7468 & 86702 & 1155 & 3934 & 791 & 9179 & 205 & 8085 \\
15579$-$5303 & 4383 & 44758 & 1956 & 6250 & 238 & 2427 & $-$ & $-$ \\
15583$-$5314 & 763 & 2824 & 594 & 1143 & 486 & 1797 & 529 & 3784 \\
16061$-$5048 & 1343 & 13273 & 422 & 1327 & 224 & 2215 & 216 & 6590 \\
16082$-$5031 & 223 & 3514 & 288 & 1144 & 189 & 2982 & 152 & 950 \\
16093$-$5015 & 1288 & 18559 & 232 & 879 & 163 & 2349 & 143 & 7815 \\
16093$-$5128 & 2709 & 5862 & 1716 & 2525 & 679 & 1470 & 300 & 1374 \\
16148$-$5011 & 1087 & 14373 & 144 & 523 & 90 & 1178 & $-$ & $-$ \\
16187$-$4932 & 626 & 4563 & 938 & 2532 & $-$ & $-$ & $-$ & $-$ \\
16204$-$4916 & 1647 & 9171 & 3434 & 8104 & 224 & 1246 & 202 & 2623 \\
16218$-$4931 & 336 & 6154 & 558 & 2389 & 107 & 1965 & 74 & 591 \\
16232$-$4917 & 262 & 3199 & 131 & 458 & $-$ & $-$ & 115 & 910 \\
16252$-$4853 & 733 & 9103 & 221 & 780 & 125 & 1553 & 202 & 8907 \\
16344$-$4605 & 4113 & 28316 & 1627 & 4269 & 299 & 2061 & $-$ & $-$ \\
16363$-$4645 & 175 & 1088 & 337 & 1681 & $-$ & $-$ & $-$ & $-$ \\
16419$-$4602 & 1302 & 21304 & 585 & 2366 & 251 & 4097 & $-$ & $-$ \\
16464$-$4359 & 2806 & 11828 & 1111 & 2281 & 526 & 2228 & 199 & 1732 \\
16501$-$4314 & 5581 & 11196 & 2324 & 3291 & 1756 & 3522 & 2798 & 7969 \\
16535$-$4300 & 2937 & 5559 & 585 & 805 & 983 & 1861 & 2140 & 5639 \\
17040$-$3959 &  & 20661 &  & 3142 &  & 2628 &  & 1100 \\
17082$-$4114 &    & 3597 &  & 800 &  & 740 &   & $-$ \\
17211$-$3537 & 2469 & 7160 & 1325 & 2257 & 399 & 1158 & 1723 & 8505 \\
17218$-$3704 &    & 820 &   & 196 &   & 180 &   & 150 \\
17256$-$3631 &    & 889 &   & 126 &  & 238 &  & 348 \\
17285$-$3346 & 210 & 2608 & 345 & 1213 & $-$ & $-$ & $-$ & $-$ \\
17338$-$3044 & 175 & 1857 & 241 & 786 & $-$ & $-$ & $-$ & $-$ \\
17355$-$3241 &    & 101 &   & 193 &   & 63 &   & 54 \\
17377$-$3109 & 78 & 29006 & 180 & 3454 & 23 & 8496 & 79 & 5914 \\
\hline
\end{tabular}
\end{center} 
\end{table*}

\end{document}